\documentclass[lettersize,journal]{IEEEtran}
\usepackage{cite}
\ifCLASSINFOpdf
\else
\fi 
\usepackage{multirow}
\usepackage{flushend}
\usepackage{graphicx}
\usepackage{colortbl}
\usepackage{caption}
\usepackage{capt-of}
\usepackage{balance}
\usepackage{subfigure}
\usepackage{fixltx2e}
\usepackage{setspace}
\usepackage{amsmath}
\usepackage{breqn}
\usepackage{hyperref}
\usepackage{array}
\usepackage[table,x11names]{xcolor}

\begin{document}
\title{Advancing Brain-Computer Interface System Performance in Hand Trajectory Estimation with \textit{NeuroKinect}} \author{ Sidharth~Pancholi*,~\IEEEmembership{Member,~IEEE,} and Amita~Giri*,~\IEEEmembership{Member,~IEEE,}
\thanks{S. Pancholi is with the Weldone School of Biomedical Engineering, Purdue University, USA (e-mail: sid.2592@gmail.com)}
\thanks{A. Giri is with the McGovern Institute for Brain Research, Massachusetts Institute of Technology (MIT), Cambridge, USA (e-mail: amita3gb@mit.edu)}
\thanks{*Authors contributed equally to this work and jointly serve as the Corresponding Authors.}
}
\maketitle
\begin{abstract}
Brain-computer interface (BCI) technology enables direct communication between the brain and external devices, allowing individuals to control their environment using brain signals. However, existing BCI approaches face three critical challenges that hinder their practicality and effectiveness: a) time-consuming preprocessing algorithms, b) inappropriate loss function utilization, and c) less intuitive hyperparameter settings. To address these limitations, we present \textit{NeuroKinect}, an innovative deep-learning model for accurate reconstruction of hand kinematics using electroencephalography (EEG) signals. \textit{NeuroKinect} model is trained on the Grasp and Lift (GAL) tasks data with minimal preprocessing pipelines, subsequently improving the computational efficiency. A notable improvement introduced by \textit{NeuroKinect} is the utilization of a novel loss function, denoted as $\mathcal{L}_{\text{Stat}}$. This loss function addresses the discrepancy between correlation and mean square error in hand kinematics prediction. Furthermore, our study emphasizes the scientific intuition behind parameter selection to enhance accuracy. We analyze the spatial and temporal dynamics of the motor movement task by employing event-related potential and brain source localization (BSL) results. This approach provides valuable insights into the optimal parameter selection, improving the overall performance and accuracy of the \textit{NeuroKinect} model. Our model demonstrates strong correlations between predicted and actual hand movements, with mean Pearson correlation coefficients of 0.92 ($\pm$0.015), 0.93 ($\pm$0.019), and 0.83 ($\pm$0.018) for the X, Y, and Z dimensions. The precision of \textit{NeuroKinect} is evidenced by low mean squared errors (MSE) of 0.016 ($\pm$0.001), 0.015 ($\pm$0.002), and 0.017 ($\pm$0.005) for the X, Y, and Z dimensions, respectively. Overall, the results demonstrate unprecedented accuracy and real-time translation capability, making \textit{NeuroKinect} a significant advancement in the field of BCI for predicting hand kinematics from brain signals.
\end{abstract}
	
	
\maketitle

\section{Introduction}
Brain-Computer Interfaces (BCIs) have revolutionized human-machine interaction by establishing a direct communication channel between the brain and external devices \cite{makin2023neurocognitive}. Various neurophysiological techniques, including Electroencephalography (EEG), Magnetoencephalography (MEG), functional Magnetic Resonance Imaging (fMRI), and functional Near-Infrared Spectroscopy (fNIRS), are employed to capture and analyze brain activity in BCIs \cite{makin2023neurocognitive, bhagat2016design}. Out of all these techniques, EEG in particular has gained maximum popularity due to its high temporal resolution, non-invasiveness, portability, and cost-effectiveness \cite{bhagat2016design}. EEG-based BCIs hold tremendous potential in developing assistive technologies that provide individuals with motor disabilities newfound independence and control over their environment \cite{zhang2019eeg, he2015wireless,giri2021cortical,gao2020classification, li2019eeg}. Collaboration among neuroscientists, computer scientists, and engineers is essential for the successful development of reliable EEG-based BCIs.

In the context of EEG-based BCIs, the literature proposed the regression-based Motion Trajectory Prediction (MTP) paradigm using multivariate linear regression (mLR) techniques. Notably, the power spectral density (PSD) of EEG signals, specifically in the delta, theta, alpha, and beta frequency bands, has emerged as a prominent feature of MTP. Several studies have explored the effectiveness of PSD-based techniques in decoding different types of movements. For instance, Robinson et al. \cite{robinson2015adaptive} employed a Kalman filter-based mLR model to decode 2D hand movements, achieving a mean correlation of $0.60 \pm 0.07$ between predicted and measured trajectories. Korik et al. \cite{korik2018decoding} demonstrated that PSD-based techniques improved the accuracy of decoding 3D imagined hand movements compared to traditional Potentials Time-Series (PTS) input. Recent work by Sosnik et al. \cite{sosnik2020reconstruction} showcased the capability of the mLR model to predict 3D trajectories of all arm joints using scalp EEG signals for both actual and imagined movements. These findings underscore the potential of PSD-based regression techniques in accurately decoding and predicting motor intentions from EEG signals. A bimanual BCI was developed to decode coordinated movements from EEG signals, achieving a correlation coefficient of 0.54 for position and 0.42 for velocity decoding \cite{chen2022continuous}. Researchers have addressed the issue of instability in neural recordings by developing stabilizers for BCIs \cite{degenhart2020stabilization}. These stabilizers align low-dimensional neural manifolds, stabilizing neural activity and preserving BCI performance. The application of stabilizers shows promise in enhancing the clinical viability of BCIs and holds potential for broader applications in other neural interfaces. 

Another important aspect of BCI research is to obtain information about the underlying neural activity and patterns associated with motor movements. This knowledge can greatly enhance our understanding of the brain's functioning and pave the way for more effective BCI systems. To achieve this, two valuable techniques, Event-Related Potentials (ERPs) for studying temporal dynamics and brain source localization (BSL) for examining spatial dynamics are useful \cite{giri2021cortical}.

By analyzing ERP waveforms, insights into neural processes in time can be gained. BSL methods identify precise neural source locations. BSL faces challenges due to factors like head geometry, conductivity, sensor noise, and spatial sampling. Two main approaches for BSL are dipole-fitting (ECD model) and dipole imaging (distributed source model). In the dipole-fitting method, a limited number of active brain regions are considered, and modeled as equivalent current dipoles (ECD). It may be noted that the number of sources is unknown prior and can be estimated by several methods proposed in the literature \cite{grech2008review,akaike1974new,wax1985detection,green1988transformation,giri2023f}. Dipole fitting involves solving an overdetermined BSL problem using nonlinear optimization techniques. Methods for dipole fitting include MUSIC \cite{mosher_multiple_1992}, RAP-MUSIC \cite{Mosher1999}, Truncated RAP-MUSIC \cite{makela_truncated_2018}, RAP Beamformer \cite{ilmoniemi2019brain}, HSH-MUSIC \cite{giri2018eeg}, H$^2$-MUSIC \cite{giri2019head,giri2020brain,giri2022anatomical}, and DS-MUSIC \cite{ilmoniemi2019brain}. The distributed source model estimates a density map of active dipoles across the cortex and employs linear optimization techniques. Common methods include MNE \cite{hamalainen1994interpreting}, dSPM \cite{DSPM}, and sLORETA \cite{pascual2002standardized}.

Deep learning emerged as a remarkable breakthrough in artificial intelligence, revolutionizing various fields, including human-computer interaction. Among its most notable applications is the accurate interpretation and decoding of brain signals, enabling the development of cutting-edge BCI. These AI-driven BCIs have opened new frontiers in neuroscience and enabled direct communication between the human brain and external devices \cite{pancholi2019improved, pancholi2019electromyography}. Deep learning techniques, such as Convolutional Neural Networks (CNNs) \cite{pancholi2022dlpr, bang2021spatio}, Long Short-Term Memory (LSTM) networks \cite{jeong2020brain},  Deep Belief Networks (DBNs) \cite{yin2017cross}, Transformers \cite{zhang2023vit}, and Autoencoders \cite{zhang2020expression}, have shown great potential in improving classification accuracy and decoding performance in EEG-based BCIs \cite{ju2022tensor, gong2021deep, khare2020time}. One notable example is the Tensor-CSPNet framework, which leverages geometric deep learning to achieve state-of-the-art performance in motor imagery EEG classification \cite{ju2022tensor}. By characterizing spatial covariance matrices on symmetric positive definite (SPD) manifolds, Tensor-CSPNet offers superior visualization and interpretability capabilities for MI-EEG classification. In the domain of motor imagery tasks, a deep neural network model has been proposed for binary-class classification using EEG signals \cite{tiwari2022midnn}. By extracting spectral features from six frequency sub-bands, the model achieves an accuracy of 72.51\% on the BCI dataset. Notably, the model's performance significantly improves to 82.48\% when utilizing band power features. This demonstrates the effectiveness of machine learning approaches in extracting meaningful information from EEG signals for accurate classification. Another noteworthy deep learning model, NeuroGrasp \cite{cho2021neurograsp}, successfully decodes multiple hands grasping movements from EEG signals. The model exhibits stable classification performance both offline and online, with an average accuracy of 0.68 ($\pm$0.09) for four-grasp-type classifications and 0.86 ($\pm$0.04) for two-grasp category classifications.

The field of BCI research has encountered various challenges, including complex pre-processing requirements, lengthy training times for deep learning models, and potential limitations in accuracy. To overcome these limitations, a novel and accurate deep learning model: \textit{NeuroKinect}, is proposed. The novelty of our work lies in the utilization of precise spatial and temporal information within a novel deep-learning architecture to learn motor task characteristics. In contrast to previous approaches that primarily focused on parameter tuning, our research emphasizes the scientific intuition behind selecting parameter values for improved accuracy. We introduce a novel loss function that considers both correlation and mean square error, setting a new standard for the accurate estimation of hand kinematics. Unlike existing studies relying on correlation analysis, our approach achieves unprecedented accuracy and real-time translation capability, making \textit{NeuroKinect} a significant advancement in predicting hand kinematics.

\begin{figure*}[t]
  \centering
  \includegraphics[width=1\linewidth]{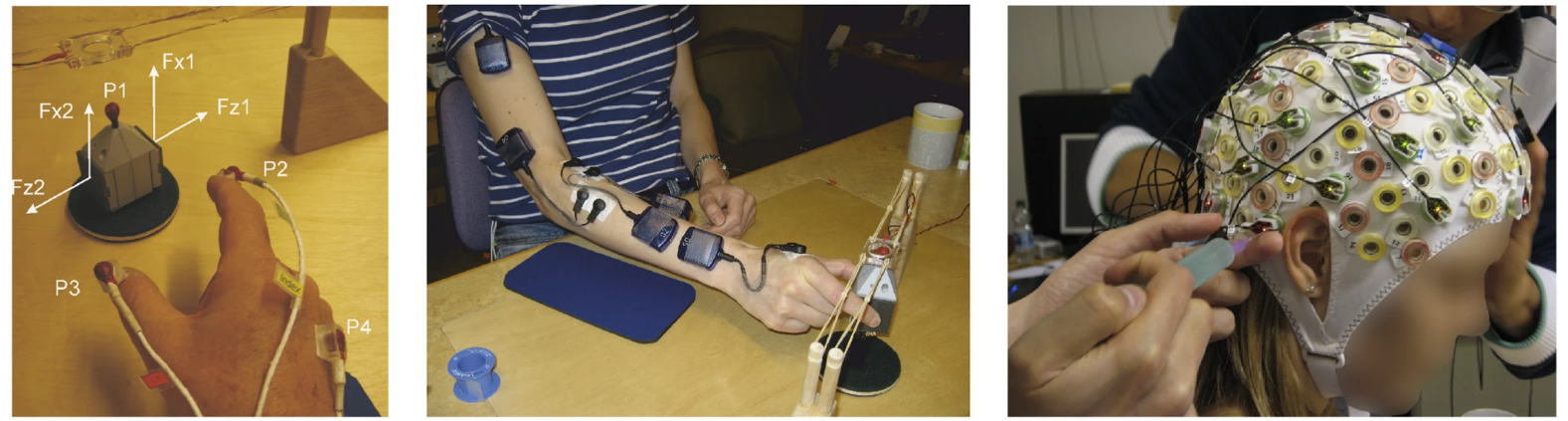}
  \caption{WAY-EEG-GAL Experimental Setup: four 3D position sensors labeled P1-P4 (left), five electromyography (EMG) sensors (middle) and 32 electroencephalographies (EEG) sensors (right) The materials used in this study are licensed under the Creative Commons Attribution 4.0 International license (CC BY 4.0), which permits unrestricted use, distribution, and reproduction in any medium, provided the original work is properly cited. \cite{luciw2014multi}}
  \label{fig:task1}
\end{figure*}

\section{EEG and Kinematic Data: Pre-processing and Data Arrangement}
\label{sec:methods}
\subsection{Data Description}
In this study, we utilize the WAY-EEG-GAL (Wearable interfaces for hAnd function recoverY-EEG-grasp and lift) dataset \cite{luciw2014multi} to investigate the relationship between the brain signals and resultant motor tasks. The dataset consists of $N=32$ channels scalp EEG recordings obtained from twelve healthy individuals while they performed grasp and lift movements using their right hand. The experiment involves a sequential set of actions: reaching, grasping, and lifting an object in a stable manner for a brief duration. Subsequently, participants were instructed to lower the object back to its initial position and return their arm to a resting state. For each subject, a total of 294 trials of the reach-to-grasp and lift task were conducted, incorporating variations in object loads and surface frictions to assess their impact on motor performance.

Figure \ref{fig:task1} provides a visual representation of the experimental setup employed for data acquisition in this study. The task initiation and the subsequent lowering of the object were prompted by an LED cue, ensuring synchronized timing across participants. To accurately capture the 3D location of the hand while performing the task, a position sensor denoted as p4 (depicted in Figure \ref{fig:task1}) was placed on the wrist of each participant. 

To gain comprehensive insights into the underlying mechanisms, simultaneous recordings of both brain signals (using the EEG sensor) and motor task signals (using the p4 position sensor) were recorded. This approach enables us to investigate the brain signals and their corresponding resultant motor tasks, facilitating the decoding of the neural mechanisms associated with motor control. Such investigations have the potential to provide valuable assistance to individuals with amputations or motor impairments, aiding in the development of advanced assistive technologies and neuroprosthetic devices.

\subsection{Kinematic Data Pre-Processing} \label{sec:kinprepro}
In a trial, 3D kinematic data was taken from the hand movement start to the hand movement stop. The p4 position sensor provides precise hand position measurements in 3D. The data was scaled and translated for each axis ($AX \in X, Y, Z$) to be within the range of 0 to 1. The process of scaling is as follows:
\begin{align}
    AX_{std} &= (AX - AX_{min}) / (AX_{max} - AX_{min}) \\
AX_{scaled} &= AX_{std} * (AX_{max} - AX_{min}) + AX_{min}
\end{align}
where $AX_{min}$ and $AX_{max}$ represents the minimum and maximum value of data in $AX$ axis. Scaling the target value is a good practice in regression modelling as it ensures that all features contribute equally to the learning process of the model.

\subsection{EEG Data Pre-Processing} \label{sec:eegprepro}
In a trial, EEG data was taken starting from the LED cue (0 ms) to the hand movement stop. The EEG and kinematic data length was made equal by removing the EEG samples at the end of each trial. To ensure compliance with our frequency of interest (0.5 Hz to 12 Hz) and the Nyquist sampling rate, we downsampled each trial for every subject from 500 Hz to 25 Hz. This downsampling process enables us to maintain the relevant frequency information within our range of interest while avoiding aliasing effects. For each EEG trial, the removal of the average voltage level, referred to as the DC component or offset, was performed on each channel by subtracting the mean value from its corresponding signal. By eliminating the DC offset, the signal's baseline is effectively adjusted to zero, facilitating a focus on the dynamic changes and fluctuations in the EEG. Following this, the EEG data underwent filtering within the frequency range of 0.5 to 12 Hz, employing a stopband attenuation of 60 dB.  

An important strength of our research lies in the omission of computationally expensive techniques like Independent component analysis (ICA) for EEG preprocessing and channel selection. This design choice ensures our approach is well-suited for real-time applications.

\subsection{Bad Trial Rejection} \label{sec:BTR}
Bad trial rejection was implemented to identify and remove trials that lacked event-related potentials and exhibited response times exceeding 500ms. These observations suggested a reduced amount of neural activity, potentially attributable to lower participant concentration during specific trials. To effectively identify and exclude these trials, a methodology was proposed, which involved the following sequential steps:

\begin{itemize}
\item{\textbf{Step 1: Maximum Moving Average}} \\
We calculated the moving average of the absolute values of the EEG data using a sliding window of length 5. In particular, we focused on the EEG data collected within a time window of 1500ms following the LED onset. This time window was selected based on our findings indicating the ERP component is typically observed within the first 1500ms, as illustrated in Figure~\ref{fig:ERPzoomed}. For each trial, the maximum moving average value was stored for each channel.\\

\item{\textbf{Step 2: Reference Signal}}\\
A reference signal, with dimensionality equal to the number of channels, was obtained by calculating the mean of the maximum moving average values across the first 10 trials. These initial 10 trials were specifically chosen due to the belief that the concentration level of participants tends to be higher at the beginning of the experiment. Furthermore, considering real-time implementation constraints, it is not practical to utilize all the available trials.

\item{\textbf{Step 3: Bad Trial Identification}}\\
Trials with a response time exceeding 500ms were initially marked as bad. Additionally, we utilized the root mean square error ($\text{RMSE}$) between the maximum moving average of remaining trials and the reference signal. If the calculated RMSE value exceeded a certain empirically determined threshold (100 and 150), the trial was flagged as bad. This threshold was set to capture significant deviations from the reference signal and indicate trials with potentially erroneous or inadequate information.
\end{itemize}

\subsection{EEG signals Standardization} 
To facilitate meaningful comparisons and analysis, the EEG signals were standardized. This involved subtracting the mean ($\mu_n$) of the EEG voltage $\nu_n$, and dividing it by the standard deviation ($\sigma_n$). Mathematically, the standardization formula can be expressed as:

\begin{equation}
E_n[t] = \frac{{(\nu_n[t] - \mu_n)}}{{\sigma_n}}
\end{equation}

where $E_n[t]$ represents the standardized EEG voltage at time $t$ and sensor $n$, and $\nu_n[t]$ represents the original EEG voltage at that time and sensor. Standardizing the EEG signals removes scale and offset differences between sensors, enhancing the reliability of comparisons and facilitating subsequent processing steps such as feature extraction or classification algorithms.

\subsection{Data Preparation } \label{sec:prep}
After ensuring the quality of EEG data through pre-processing, bad trial rejection and signals standardization, the subsequent step involves preparing the data for input into the proposed model. 

A key consideration in this data preparation process was addressing the temporal differences between the EEG and kinematic data, arising from the inherent delay in \textit{"processing"} and \textit{"transferring"} neural information from the brain to motor control. To address this, as outlined in Section \ref{sec:kinprepro} and \ref{sec:eegprepro}, length of EEG and kinematic data were made equal by removing the EEG samples from the end of trial. This adjustment allows us to capture the pre-movement neural information that contributes to subsequent motor activity. By aligning the temporal dimensions of the datasets, we can effectively analyze the relationship between neural signals and motor control. Following this step, the EEG and kinematic data trials of cell structure were transformed into matrices using the "cell2mat" command in MATLAB. It maybe noted that the resultant EEG data is of dimension $N \times T$, where $N$ represents the number of channels, and $T$ represents the time. On the other hand, the resultant kinematic data is of dimension $3 \times T$, representing the hand's kinematic trajectory in the $x$, $y$, and $z$ directions. Our objective is to utilize the EEG signals to predict the hand kinematics. 

To enhance the probability of EEG segment corresponding to the observed kinematic data, "\textbf{Input}" dat was prepared that contains the delayed version of original EEG and is defined as
\begin{align}
  \textbf{Input}(t+l) &= \text{Conc} [k_0, k_1, \cdots k_l ]  \\
  k_i &= E_n[t+l-i]~~~~~~i\in[0,l]
\end{align}
where "conc" is the concatenation of EEG segments and $l$ represents the EEG lags. Dimension of "\textbf{Input}" is $N*(l+1) \times (T-l)$. The choice of EEG lags was determined based on our analysis of temporal and spatial dynamics, as detailed in Section \ref{tsd}. Additionally, a delay $d$) was incorporated to account for the transfer delay.

In summary, the "\textbf{Input}($t+l$)" representation was fed into the model to estimate the 3D position of the hand kinematics at time $(t+l+d)$. By incorporating this delayed and concatenated EEG data, we aimed to improve the alignment between the observed kinematic data and the corresponding EEG segments, facilitating more accurate prediction of hand kinematics within our model. In contrast to previous approaches that primarily focused on these parameter tuning, our research emphasizes the scientific intuition behind selecting parameter values, detailed in Section \ref{tsd} and \ref{sec:prostages}.

\section{Proposed Deep Learning Model: \textit{NeuroKinect} }
\begin{figure*}[htbp]
  \centering
  \includegraphics[width=1.2\linewidth]{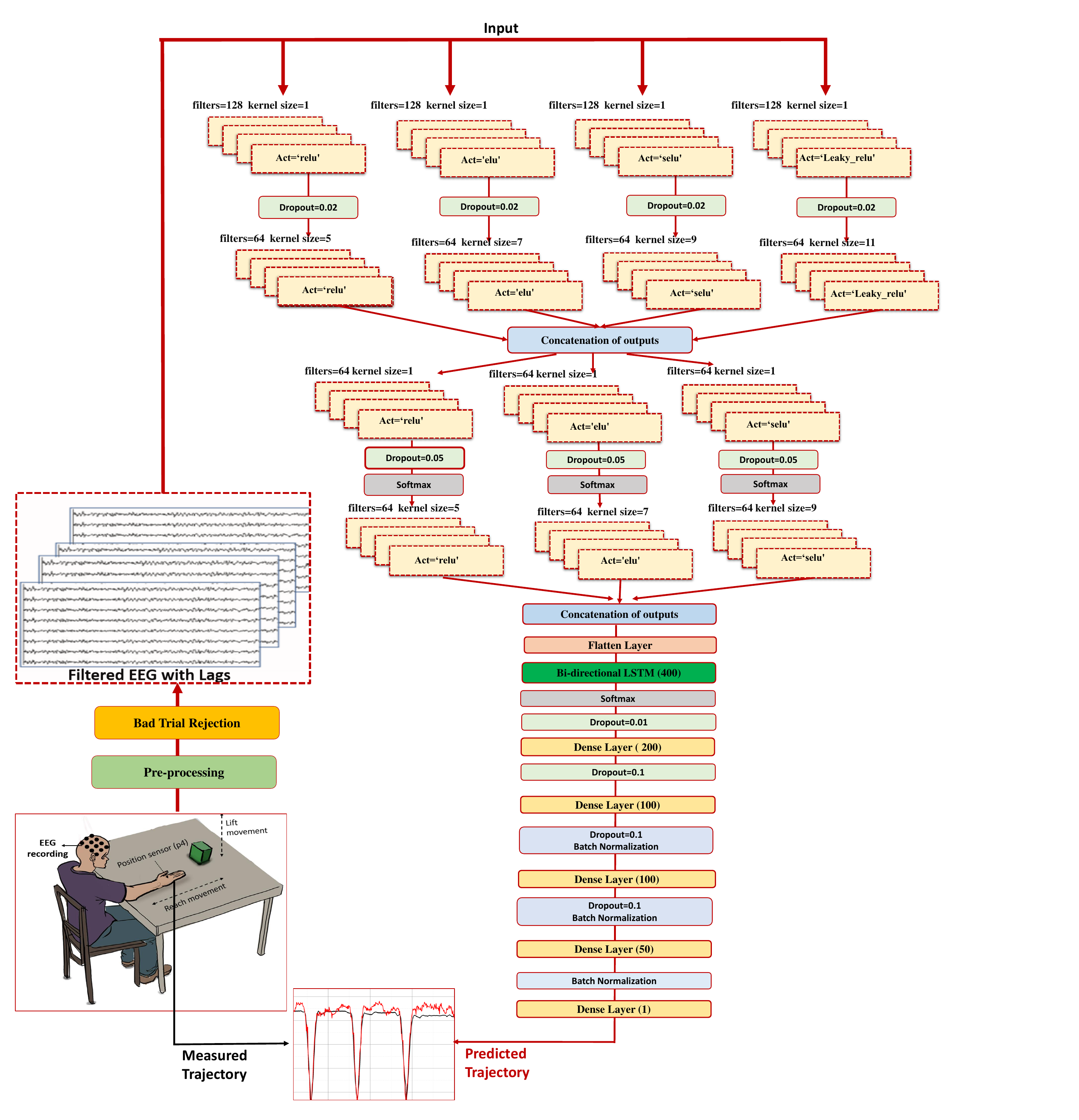}
  \caption{Architecture of the proposed deep learning model: \textit{NeuroKinect}}
  \label{fig: NeuroKinect}
\end{figure*}

The proposed deep learning model NeuroKinect is specifically designed to predict hand kinematics using EEG signals and is presented in Fig \ref{fig: NeuroKinect}. It comprises a stacked architecture with time-distributed convolutional, recurrent, and dense layers. One of the key innovations of this model is the utilization of 1D convolutional layers with a kernel size of 1, which is inspired by google's inception network  \cite{szegedy2015going}. This design choice offers several advantages. Firstly, the use of a kernel size of 1 allows the model to capture fine-grained patterns and learn intricate relationships between features. By focusing on local information, the model becomes more sensitive to subtle variations in the EEG signals that are indicative of hand movements. Furthermore, this approach facilitates dimensionality reduction. By applying convolutions with a kernel size of 1, the spatial dimensions are reduced while preserving important information. This helps to condense the input data without losing relevant features, leading to a more compact representation. Moreover, the adoption of smaller kernel sizes enhances computational efficiency, particularly in deep architectures and with large-scale datasets. The smaller kernel size reduces the number of parameters and computational operations required, resulting in faster training and inference times. To introduce non-linearity and capture diverse information from the EEG signals, the model incorporates various activation functions such as ReLU, ELU, SELU, and Leaky ReLU. These activation functions enhance the model's ability to learn complex relationships within the EEG data and accurately reconstruct the hand kinematics.

In order to optimize the model's performance, the input sequence is first passed through a Bidirectional LSTM layer, a type of recurrent neural network that processes the sequence in both forward and backward directions. This allows the model to leverage information from both past and future contexts, enhancing its training capabilities. Following the Bidirectional LSTM layer, a series of dense layers is applied. These layers incorporate nonlinear activation functions and regularization techniques, such as dropout and batch normalization, in order to prevent overfitting and improve the model's ability to generalize. To further enhance the model's performance, softmax activation is utilized within the CNN blocks and after the Bidirectional LSTM layer. This activation function provides benefits such as output normalization, capturing complex feature interactions, aiding in gradient propagation, and fostering ensemble-like behavior \cite{totaro2020non}. These aspects contribute to the model's ability to effectively capture diverse aspects of the data. During the training process, a custom callback is employed to evaluate correlation and MSE metrics on a validation set after each epoch. By comparing the current correlation and MSE values with the previous best values, the callback determines if the model should be saved. This approach ensures that the saved model achieves the highest possible correlation while also considering the MSE, optimizing the training process for accurate hand kinematics reconstruction.

To balance the computational efficiency and model performance, a batch size of 100 is chosen for training. This means that the model processes 100 samples at a time before updating its parameters. By leveraging parallel computation capabilities, this approach enables faster training without sacrificing the accuracy of the model.

For the task-specific optimization of model parameters, a customized loss function ($\mathcal{L}_{\text{Stat}}$ ) was employed and detailed in the following section. The Adam optimizer, a widely used optimization algorithm in deep learning, is then utilized to update the model's weights and biases based on the gradients of the loss function. The adaptive learning rate provided by the Adam optimizer facilitates efficient convergence during training, further optimizing the model's performance.

\subsection{Proposed Loss Function: ${\mathcal{L}_{\text{Stat}}}$} \label{sec:losspro}

The proposed loss function aims to enhance the predictive performance of the \textit{NeuroKinect} model by maximizing the Pearson correlation coefficient ($\rho$) (as in eqn. \ref{corr}) while ensuring similar statistical parameters, such as mean and variance, between the predicted kinematics ($x$) and the target kinematics ($y$). The loss function, denoted as $\mathcal{L}_{\text{Stat}}$, is defined as follows:

\begin{dmath}
\mathcal{L}_{\text{Stat}} = \underbrace{1 - \rho}_{\substack{\text{1st term}}} + \underbrace{0.01 \left( \frac{\sum_{i=1}^{n} (x_i - y_i)^2}{\sqrt{n} \sqrt{\sum_{i=1}^{n} (y_i - \bar{y})^2}} \right)}_{\substack{\text{2nd term}}} + \underbrace{0.1 \left| \frac{\left(\sum_{i=1}^{n} (x_i - \bar{x})^2\right) - \sum_{i=1}^{n} (y_i - \bar{y})^2}{\sum_{i=1}^{n} (y_i - \bar{y})^2} \right|}_{\substack{\text{3rd term}}}
\label{loss}
\end{dmath}

The proposed loss function comprises three components. The first term encourages a higher correlation between predicted and target kinematics. The second term penalizes the squared difference between predicted and target kinematics, normalized by the square root of the sum of squared differences of the target kinematics from their mean. The third term penalizes the absolute difference between the difference of squared differences of the predicted and target kinematics' means and the sum of squared differences of the target kinematics from their mean. These terms collectively optimize the model by promoting correlation, controlling mean squared error, and ensuring similarity in variance between predicted and target kinematics.

\subsection{Metrics for Performance Evaluation}

In evaluating the performance of our proposed model for trajectory estimation, we utilized two widely employed metrics: the Pearson Correlation Coefficient $\rho$ and MSE. The Pearson correlation coefficient quantifies the strength and direction of the linear relationship between two variables $x$ and $y$ and is defined as:
\begin{equation}
    \rho = \frac{{\sum_{i=1}^{n}(x_i - \bar{x})(y_i - \bar{y})}}{{\sqrt{{\sum_{i=1}^{n}(x_i - \bar{x})^2} \sum_{i=1}^{n}(y_i - \bar{y})^2}}}.
    \label{corr}
\end{equation}
where $n$ is the number of data points along the specified axis. The correlation coefficient $\rho$ ranges between -1 and 1, where a value of 1 or -1 indicates a strong positive or negative linear relationship, respectively. Conversely, a value closer to 0 suggests a weaker or no linear relationship between the variables. For evaluating 3D trajectories, a more natural 3D correlation coefficient (${\rho}_{3D}$) is defined, which calculates the average of the $\rho$ values obtained independently for each axis ${\rho}_x$, ${\rho}_y$, ${\rho}_z$.  

\begin{equation}
{\rho}_{3D} = \frac{{{\rho}_x + {\rho}_y + {\rho}_z}}{3}
\end{equation}

On the other hand, the MSE calculates the average squared difference between the predicted trajectory values, $x$, and the true trajectory values, $y$. The summation is taken over $n$ data points, and the resulting sum is divided by $n$ to obtain the mean squared difference. The MSE provides an overarching measure of prediction error, with lower values indicating enhanced accuracy and precision in trajectory estimation. 

For evaluating 3D trajectories, we utilize 3D MSE ($\text{MSE}_{3D}$), which measures the average of MSE values obtained independently for each axis $\text{MSE}_x$, $\text{MSE}_y$, $\text{MSE}_z$. 

\begin{equation}
\text{MSE} = \frac{\sum_{i=1}^{n} (x_i - y_i)^2}{n}
\end{equation}

\begin{equation}
\text{MSE}_{3D} = \frac{{\text{MSE}_x + \text{MSE}_y + \text{MSE}_z}}{3}
\end{equation}

\section{Neural Decoding of Hand Kinematics in Time and space} \label{tsd}
Temporal and spatial information are crucial factors that significantly influence the overall performance of the model, as they guide us in providing the relevant information necessary for the model to learn the underlying characteristics of the process. In this study, we utilize two approaches to investigate these aspects: Event-Related Potentials (ERPs) for studying temporal dynamics and brain source localization for examining spatial dynamics. By analyzing the ERP waveform and examining the associated brain activity regions, valuable insights into the underlying neural processes can be obtained. Detailed analysis of the temporal and spatial brain activity underlying current task is presented in the next subsequent section.

To ensure dataset reliability, trials with a response time delay exceeding 500 ms between the LED cue and movement were excluded. Among the 294 total trials per subject, Table \ref{tab:trials} summarizes the remaining trials (column 2) after applying the exclusion criteria. It is worth noting that subjects whose remaining trials accounted for less than 60\% of the total trials were excluded from further analysis due to insufficient data. Consequently, Participant 2, 5, 7, and 12 were excluded from the study. For certain subjects, Fp1 and Fp2 channels exhibited significant levels of noise, likely resulting from their detachment from the scalp while data recording. Consequently, these channels were identified as "bad" and were excluded from the current analysis of spatial and temporal dynamics. In addition, we provide key statistical measures of the response time in columns 3 and 4 of the table. The mean response time (MRT) represents the average time taken by the participants to initiate the motor task following the LED cue, while the standard deviation of response time (SDRT) indicates the variability in response times across the trials. These measures contribute to our understanding of the temporal and spatial dynamics of motor responses observed in the study. 

\begin{table}
\centering
 \caption{Participant-wise information for the selected number of trials (1st column) out of a total 294, based on response time less than 500ms. MRT and SDRT represent the mean and standard deviation of response time for selected trials. }
\begin{tabular}{ccccc}

    \hline
    \textbf{Participant} & \textbf{No. of Trials} & \textbf{MRT} & \textbf{SDRT} \\
    \hline
    P1  & 284 & 0.33 & 0.06  \\
    \rowcolor{lightgray} P2  & 19  & -    & -   \\
    P3  & 293 & 0.32 & 0.04 \\
    P4  & 287 & 0.32 & 0.05 \\
    \rowcolor{lightgray}P5  & 120 & -    & -  \\
    P6  & 213 & 0.39 & 0.06 \\
    \rowcolor{lightgray}P7  & 23  & -    & -   \\
    P8  & 263 & 0.37 & 0.06 \\
    P9  & 236 & 0.37 & 0.07 \\
    P10 & 220 & 0.38 & 0.06 \\
    P11 & 199 & 0.38 & 0.06 \\
    \rowcolor{lightgray}P12 & 154 & -    & -    \\
    Avg & 250 & 0.36 & 0.06 \\
    \hline
    \end{tabular}
   
    \label{tab:trials}
    \end{table}

\begin{figure}[t]
  \centering
  \includegraphics[width=1\linewidth]{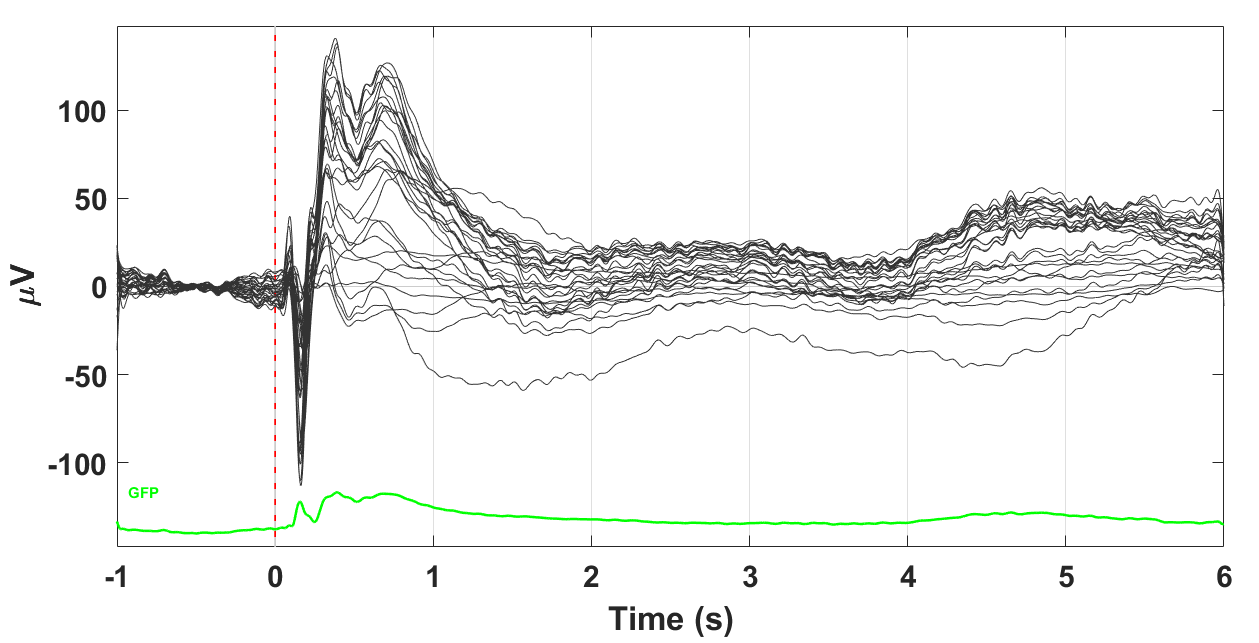}
  \caption{Grand average EEG response across subjects and trials.}
  \label{fig:ERP}
\end{figure}

\subsection{Event-Related Potentials} 

The ERP, or event-related potential, refers to the measured electrical activity of the brain in response to specific stimuli or events. It is obtained by averaging the brain's electrical responses across multiple trials to extract the time locked neural activity related to the stimulus. In our study, a trial was defined by segmenting the EEG data into epochs of 7 seconds. Each trial consisted of 1 second before the LED onset data and 6 seconds after the LED onset data. 

In \cite{pancholi2022source}, the frequency range of 0.5-12 Hz consistently yielded the best results for all the decoding methods employed. Building upon these findings, we adopted the same frequency range in the current study to filter the EEG data. To prepare the data for analysis, we applied a bandpass filter with a range of 0.5 to 12 Hz and a stopband attenuation of 60 dB to each trial per subject. This filtering step helps isolate the frequency components that are most relevant for our study. Additionally, baseline removal was performed by calculating the mean over the baseline period (0 to 1 second) and subtracted from the data. Next, we computed the average EEG response for each subject by averaging the trials. This subject-dependent average EEG response allowed us to analyze the individual subject-level data. 

For the group-level analysis, we further averaged the subject's average EEG responses to calculate the grand average EEG response. This aggregated data provided insights into the collective response pattern across subjects, enabling a comprehensive understanding of the underlying neural processes. The resulting grand average EEG response is presented in Figure~\ref{fig:ERP}, with the LED onset indicated at 0 ms. Task-related EEG locked response can be seen within the first 1.5 seconds after the LED onset. The early response likely captures the initial processing stages and may include components such as sensory processing (visual), attentional allocation, stimulus evaluation, motor planning and Response Execution. 

\begin{figure*}[t]
  \centering
  \includegraphics[width=0.8\linewidth]{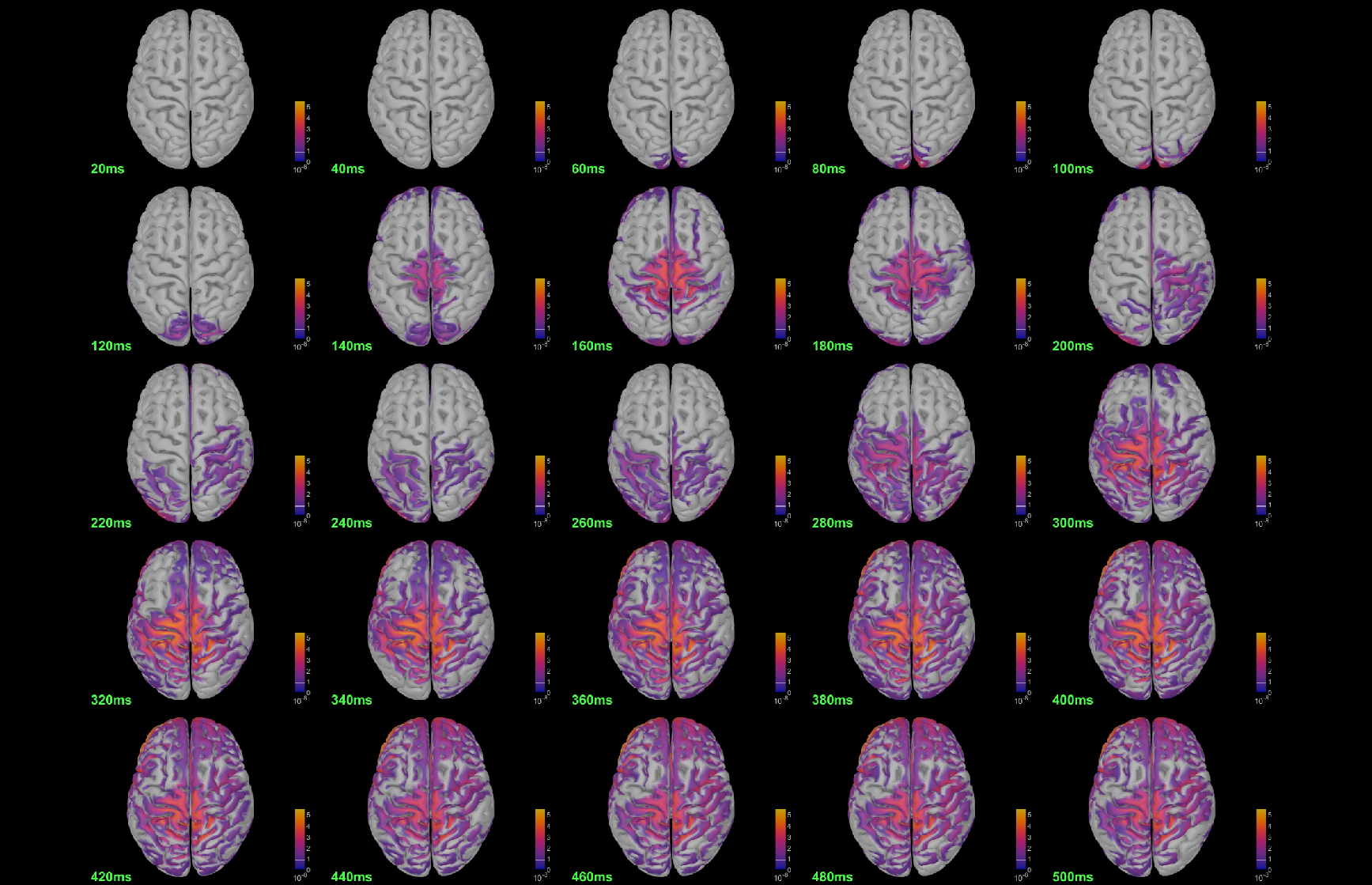}
  \caption{25 BSL plots using sLORETA at different time stamps sliced temporally from 20ms to 500ms.}
  \label{fig:sloreta}
\end{figure*}


\subsection{Brain Source Localization}
To gain a more comprehensive understanding of these components, brain source localization was performed to identify the spatial regions of the brain that are involved in processing at different time points. 


\begin{figure}[t]
  \centering
  \includegraphics[width=1\linewidth]{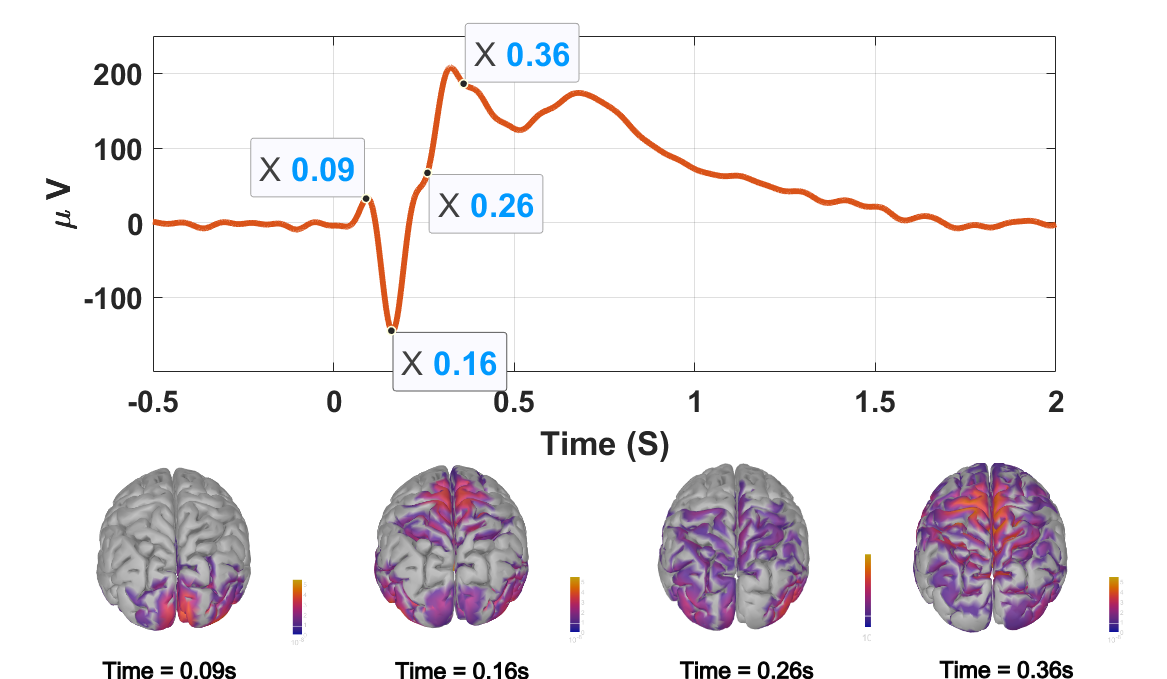}
  \caption{ERP and BSL at different time stamps.}
  \label{fig:ERPzoomed}
\end{figure}


In this particular study, the sLORETA dipole imaging method was chosen for source localization. This method utilizes images of standardized current density while considering the constraint of smoothly distributed sources for inferring localization. Source localization plots for the activity under consideration (right-hand grasp and lift (GAL) execution task) are given in Figure \ref{fig:sloreta}. A visual cue for the start of the activity was presented at 0 ms. There are a total of 25 images of cortical surface activation sliced temporally from 20ms to 500ms. The BSL result shown corresponds to the group average EEG presented in Fig.~\ref{fig:ERP}. At around 60 ms to 120 ms following the onset of the LED, a noticeable illumination emerges in the primary visual cortex, indicating the initiation of early visual processing. Between 120 ms and 260 ms, there appears to be a transfer of information from the visual cortex to the motor cortex, signifying motor planning stages. After motor planning, the prepared motor commands are transmitted to the motor cortex between 260ms to 360ms, which activates the relevant muscles for hand movement. The execution of the hand movement was observed 360ms on wards which involves a cascade of neural signals traveling from the motor cortex to the spinal cord and finally to the muscles controlling hand movement. The temporal information received from BSL results will be utilized to assign values to the parameters ($l,d$) described in Section \ref{sec:prep}. These parameters are crucial for data preparation, encompassing considerations for \textit{processing} and \textit{transferring} delays. In contrast to previous approaches that primarily focused on these parameter tuning, our research emphasizes the scientific intuition behind selecting parameter values.

In the Results section, we begin by providing a comprehensive description of each neural decoding stage relative to the onset of the LED. This temporal framework is essential for understanding the significance of these stages and their role in parameter value assignment.

\section{Results}

\subsection{Neural Processing Stages} \label{sec:prostages}
To examine both spatial and temporal dynamics simultaneously, Figure ~\ref{fig:ERPzoomed} presents the grand average event-related potential alongside the results of source localization. The ERP is obtained by summing the potential across all channels of the grand average EEG (in Figure~\ref{fig:ERP}). Next will describe each neural processing stages that provide insights into the ERP components:

\textbf{Visual Sensory processing (60ms to 120ms)}: The early positive response, referred to as P1 in ERP was observed around 60-120 ms after stimulus onset, signifying the initial processing of visual stimuli. By localizing the brain sources associated with this response, we have successfully identified the activation of the occipital cortex, situated at the posterior region of the brain, as the specific source correlated with this response. This activation serves as an indication of the occipital cortex's involvement in visual sensory processing.

\textbf{Attention allocation, Stimulus evaluation and Motor planning (120ms to 260ms)}: 
The negative deflection occurring approximately 120-260 ms after stimulus onset is linked to the categorization and evaluation of visual stimuli. Once the initial sensory processing takes place, the brain actively allocates attention to evaluate the stimuli and determine their significance or meaning. This evaluation process engages various brain regions, including the prefrontal cortex, parietal cortex, and the limbic system, which collectively contribute to the assessment of the stimuli. Subsequently, motor planning processes commence, involving higher-level cognitive processing, which may require additional time to unfold. These processes typically occur within a few hundred milliseconds after stimulus onset, as the brain prepares and organizes the necessary motor actions for the hand movement task.

During the 120ms to 260ms time frame, activation can be observed in the prefrontal, parietal, and motor cortex, indicating the allocation of attention, stimulus evaluation, and motor planning activities. This concurrent activation in these regions provides evidence of their involvement in these cognitive processes.

\textbf{Response execution (260ms to 360ms)}: 
After the completion of motor planning, the prepared motor commands are transmitted to the motor cortex, triggering the activation of the specific muscles required for hand movement. The execution of the hand movement entails a series of neural signals cascading from the motor cortex to the spinal cord and ultimately to the muscles responsible for controlling hand movements. This transmission process requires several hundred milliseconds to relay the information from the brain to the muscles.

During this period, significant activation in the motor cortex can be observed, indicating its central role in coordinating and initiating motor commands. Actual hand movement was observed at 360ms after the LED onset (please refer to average MRT of Table \ref{tab:trials}). Our analysis revealed a neural \textit{processing} duration of approximately 260ms and a transfer delay of around 100ms for kinematic muscles. Considering a sampling frequency of 25 Hz, we carefully selected EEG lags of $l=10$ to define a time window of 250ms. Transfer delay $d=4$ was utilized to account for 100ms \textit{transferring} information.

\subsubsection{Loss Function Analysis}
The advantage of utilizing the proposed loss function $\mathcal{L}_{\text{Stat}}$ (presented in Section \ref{sec:losspro}) over existing studies that rely solely on correlation analysis is highlighted in this section. 

\begin{figure*}[!t]
	\centering
	\subfigure[]{\includegraphics[width=0.31\textwidth]{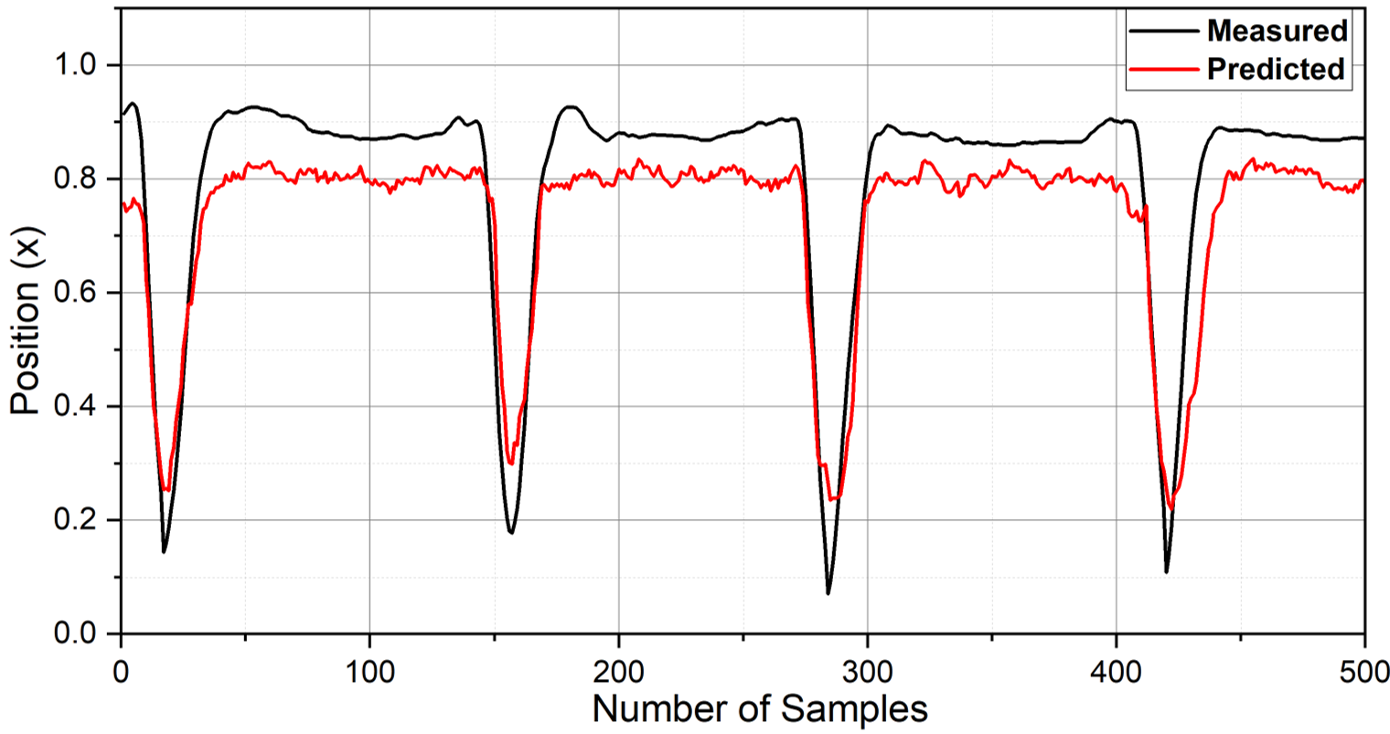}}
	\subfigure[]{\includegraphics[width=0.31\textwidth]{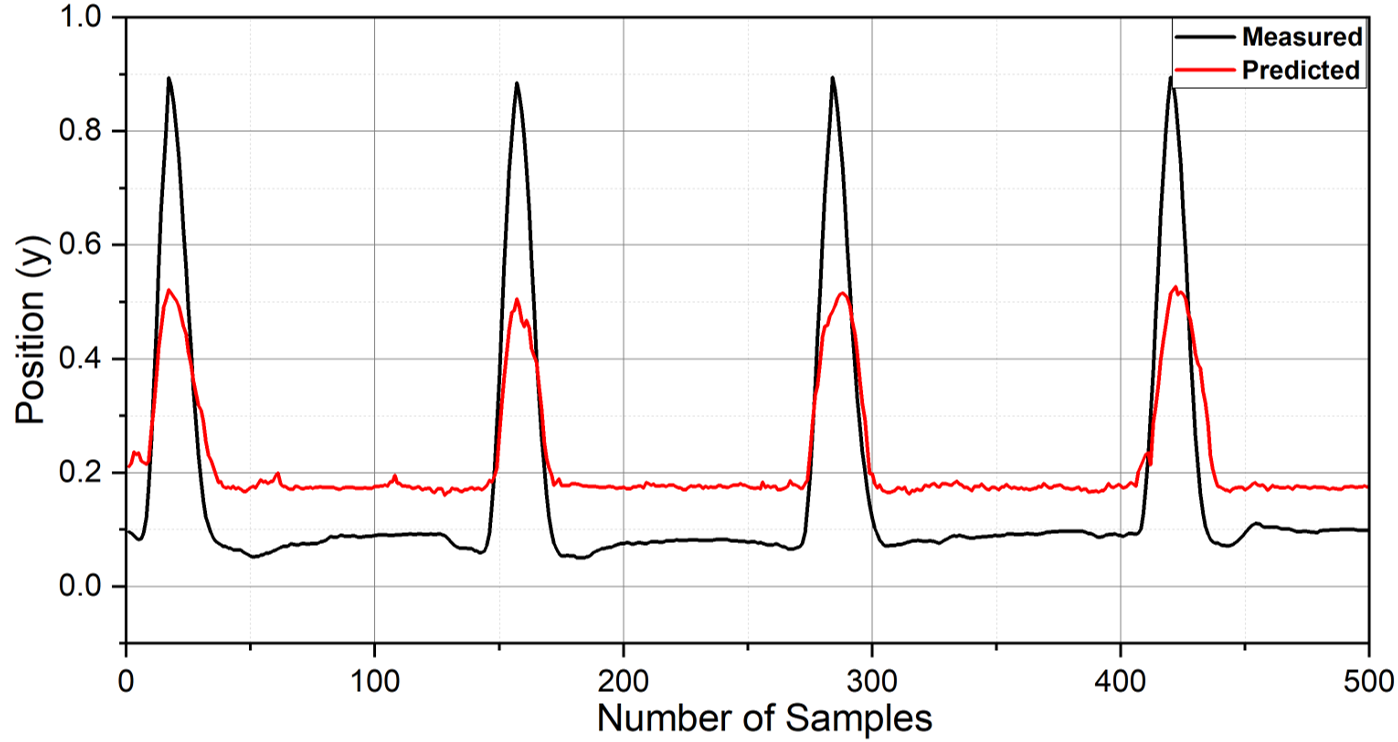}}
	\subfigure[]{\includegraphics[width=0.31\textwidth]{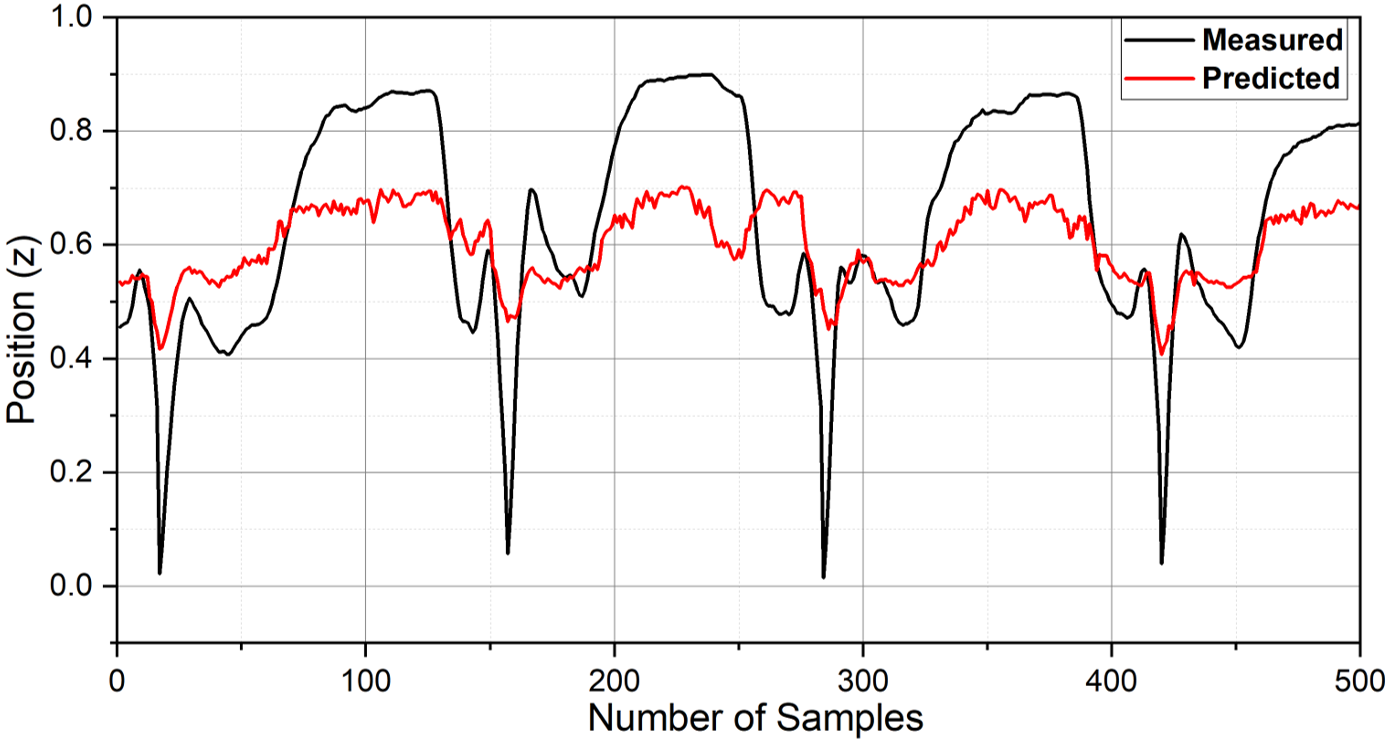}}
  \caption{
 Addressing Correlation-MSE Discrepancy in Hand Kinematics Prediction. The correlation coefficients indicate strong association, but high MSE values reveal significant deviations between predicted and measured values.}
  \label{fig:lcorr}
\end{figure*}

To address the discrepancy between correlation and mean square error in hand kinematics prediction, we conducted analysis using Subject P10 as an example. For this subject, the correlation coefficients between the predicted and measured values in the X, Y, and Z directions are found to be ${\rho}_x$ = 0.96, ${\rho}_y$ = 0.95, and ${\rho}_z$ = 0.85, indicating a high degree of association. However, the corresponding MSE values, $\text{MSE}_x$ = 0.15, $\text{MSE}_y$ = 0.023, and $\text{MSE}_z$ = 0.032, remain relatively high. This implies that despite the strong correlation, the predicted values deviate significantly from the measured values, resulting in larger mean squared errors. Figure~\ref{fig:lcorr} visually demonstrates the discrepancy between correlation and MSE. Therefore, it is crucial to employ a loss function that accounts for correlation and MSE. By considering both correlation and mean square error, $\mathcal{L}_{\text{Stat}}$ sets a new standard for accurately estimating hand kinematics as presented in next Section.

\subsection{Hand Kinematic Trajectory Estimation Analysis}
In this section, we present the analysis of hand kinematic trajectories in the X, Y, and Z directions after performing EEG and kinematic data preprocessing, as well as arranging the data with appropriate delays and window sizes. 
\subsubsection{X Direction Analysis}

\begin{figure}[h!]
  \centering
  \includegraphics[width=0.72\linewidth]{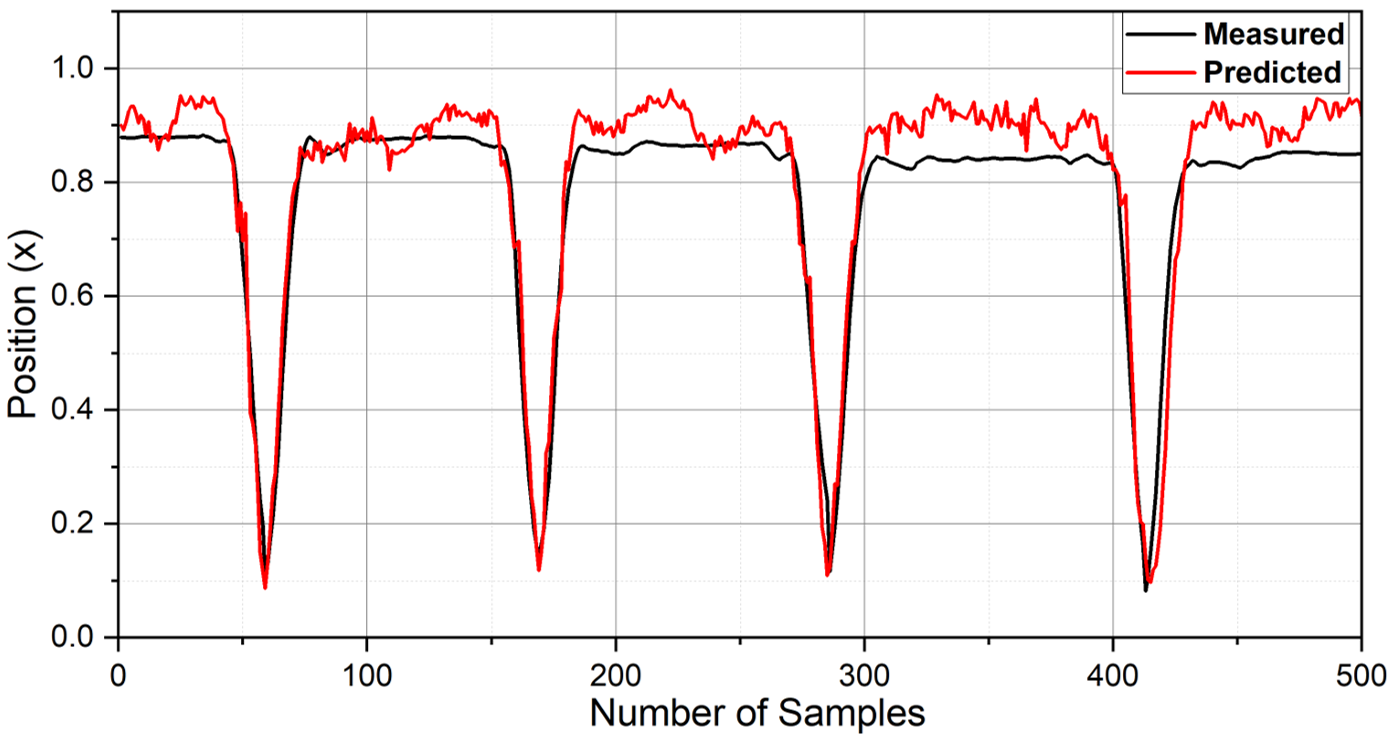}
  \caption{
Measured and Predicted Values for the x-direction.}
  \label{fig:x_result}
\end{figure}

The model demonstrated impressive performance in predicting the kinematics of the X direction, accurately capturing the complexities and variations present in the corresponding EEG signals as shown in Fig. \ref{fig:x_result}. The correlation coefficients ranged from 0.88 to 0.95, indicating a strong positive relationship between the predicted and measured values. This highlights the model's ability to effectively capture the nuances of the X direction. Performance metrics for X directional analysis is presented in Table \ref{tab:performance_metrics}.

Subject P10 achieved an outstanding correlation coefficient of 0.95 in the X direction, showcasing remarkable accuracy in predicting kinematics. On the other hand, Subject P3 displayed a slightly lower correlation coefficient of 0.88, indicating a relatively weaker but still notable association. Overall, the model consistently performed well in the X direction, with an average correlation coefficient of 0.92 and a narrow standard deviation of 0.03, demonstrating reliable performance in capturing and predicting the complexities of hand kinematics.

The MSE values further support the model's proficiency in the X direction, ranging from 0.006 to 0.020. On average, the MSE across all subjects in the X direction is 0.016, with a standard deviation of 0.006, indicating highly accurate predictions with some variability. Subject P1 achieved an exceptional MSE value of 0.006, demonstrating the model's precise prediction of the measured values for this specific subject. On the other hand, Subject P3 exhibited the highest MSE value of 0.022, indicating a relatively larger discrepancy between the predicted and measured values in the X direction.

\vspace{3cm}
\begin{table*}[t]
\caption{Performance metrics for X, Y, and Z Directions}
\centering
\label{tab:performance_metrics}
\begin{tabular}{cccccccccc}
\hline
\textbf{Participant} & ${\rho}_x$ & ${\rho}_y$ & ${\rho}_z$ & ${\rho}_{3D}$ & \text{MSE$_x$} & \text{MSE$_y$} & \text{MSE$_z$} & \text{MSE$_{3D}$} \\
\hline
P1     & 0.95 & 0.95 & 0.87 & 0.923 & 0.006 & 0.006 & 0.012 & 0.008 \\
P3     & 0.88 & 0.88 & 0.84 & 0.867 & 0.020 & 0.020 & 0.019 & 0.020 \\
P4     & 0.90 & 0.91 & 0.84 & 0.883 & 0.015 & 0.017 & 0.025 & 0.019 \\
P6     & 0.92 & 0.93 & 0.71 & 0.853 & 0.020 & 0.021 & 0.023 & 0.021 \\
P8     & 0.93 & 0.93 & 0.83 & 0.900 & 0.022 & 0.017 & 0.015 & 0.018 \\
P9     & 0.92 & 0.92 & 0.83 & 0.890 & 0.017 & 0.012 & 0.016 & 0.015 \\
P10    & 0.95 & 0.96 & 0.90 & 0.937 & 0.019 & 0.018 & 0.014 & 0.017 \\
P11    & 0.90 & 0.93 & 0.79 & 0.873 & 0.013 & 0.011 & 0.018 & 0.014 \\ \hline
\rowcolor{lightgray}Proposed (Avg) & 0.92 & 0.93 & 0.83 & 0.89 & 0.016 & 0.015 & 0.017 & 0.016 \\
\hline
\end{tabular}
\end{table*}

\begin{figure}[h!]
  \centering
  \includegraphics[width=0.72\linewidth]{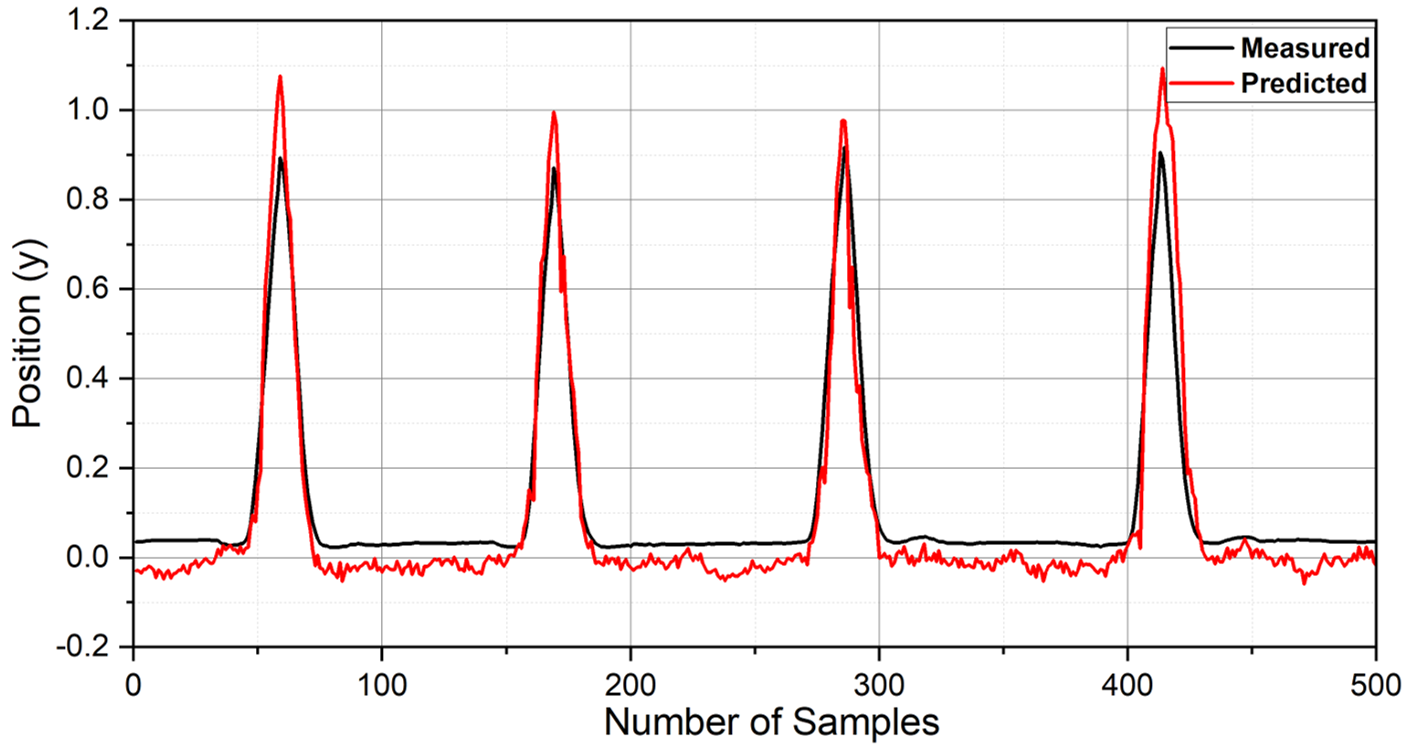}
  \caption{Measured and Predicted Values for the y-direction.}
  \label{fig:y_result}
\end{figure}

\begin{figure}[h!]
  \centering
  \includegraphics[width=0.72\linewidth]{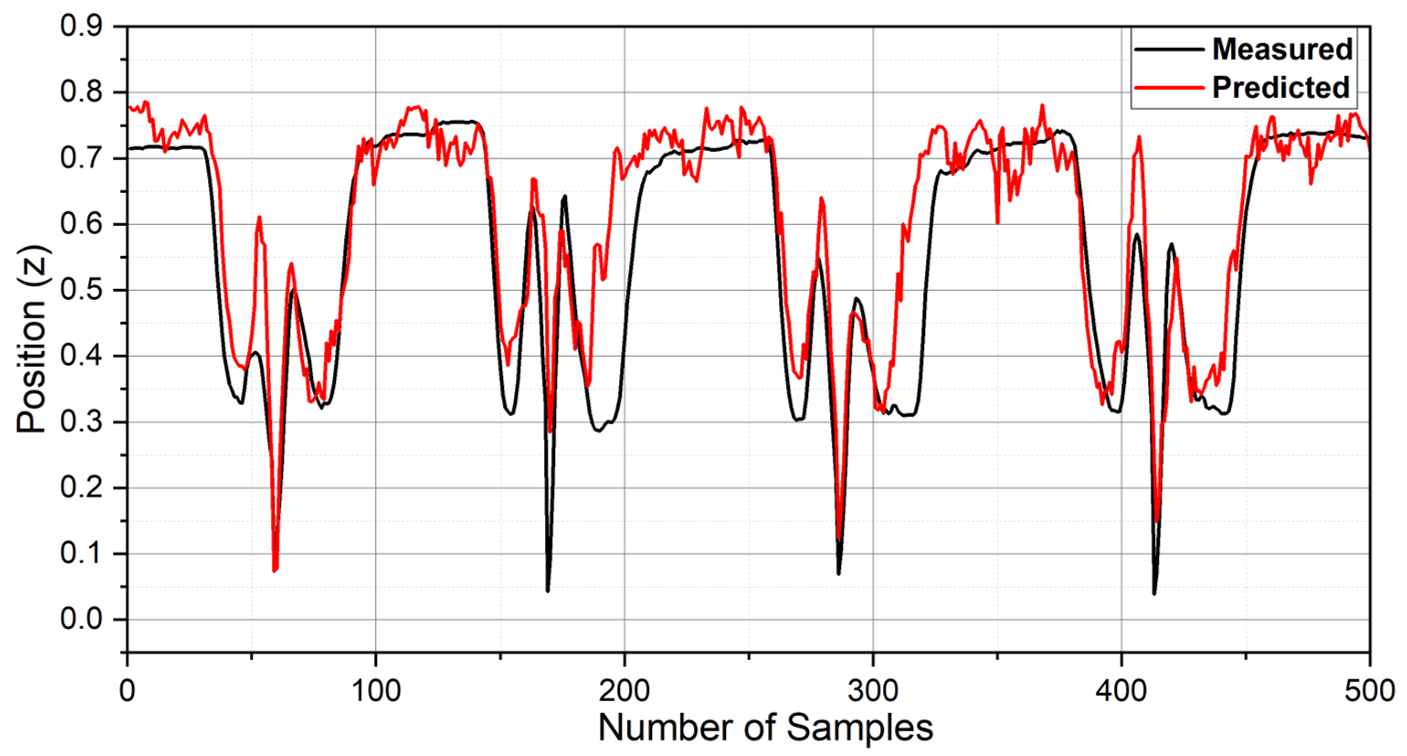}
  \caption{Measured and Predicted Values for the z-direction..}
  \label{fig:z_result}
\end{figure}

\subsubsection{Y Direction Analysis}

The model's performance in predicting the kinematics of the Y direction was remarkably accurate, highlighting its exceptional ability to anticipate hand movements based on corresponding EEG signals as showcased in Fig. \ref{fig:y_result}. Throughout the evaluation, correlation coefficients ranging from 0.88 to an impressive 0.96 consistently indicated a strong positive relationship between the predicted and measured values. Subject P10 notably achieved the highest correlation coefficient of 0.96 in the Y direction, affirming the model's exceptional accuracy in predicting hand sensor measurements for this specific subject. Conversely, Subject P3 demonstrated a slightly lower yet commendable correlation coefficient of 0.88. Performance metrics for Y directional analysis is presented in Table \ref{tab:performance_metrics}.

The MSE values in the Y direction further validate the model's accuracy, ranging from 0.011 to 0.021. These values indicate the precision with which the model predicts the measured values. The average MSE of 0.016, with a narrow standard deviation of 0.005, reflects the highly accurate predictive performance of the model. Subject P1 achieved an outstandingly low MSE value of 0.006, demonstrating the model's exceptional precision in predicting the measured values in the Y direction for this subject. Conversely, Subject P3 had the highest MSE value of 0.021, suggesting a slightly larger discrepancy between the predicted and measured values.

\subsubsection{Z Direction Analysis}

In the Z direction, the model's performance was notable, although slightly lower compared to the X and Y directions as shown in Fig.\ref{fig:z_result}. The correlation coefficients ranged from 0.7 to 0.9, indicating a moderate positive correlation between the predicted and measured values. The average correlation coefficient of 0.83, with a standard deviation of 0.06, reflects the overall relationship between the EEG signals and the Z direction. Performance metrics for Z directional analysis is presented in Table \ref{tab:performance_metrics}.

Subject P10 achieved the highest correlation coefficient of 0.9 in the Z direction, indicating a strong linear relationship between the predicted and measured values. On the other hand, Subject P6 exhibited the lowest correlation coefficient of 0.71, representing a relatively weaker association.

Considering the MSE values in the Z direction, they ranged from 0.012 to 0.025. The average MSE of 0.017, with a standard deviation of 0.004, suggests relatively accurate predictive performance. Subject P1 achieved the lowest MSE value of 0.012, indicating high precision in predicting the measured values in the Z direction. Conversely, Subject P4 had the highest MSE value of 0.025, implying a slightly larger deviation between the predicted and measured values. These results further illustrate the model's ability to provide accurate predictions in the Z direction, while accounting for some variability among different subjects.

\subsection{3D Metrics Analysis}
The model's performance in the three-dimensional space (${\rho}_{3D}$) was remarkable, showcasing its ability to capture and interpret the complex relationships between the EEG signals and the hand sensor measurements across all three directions. With an average correlation coefficient of 0.89 and a narrow standard deviation of 0.03, the model consistently demonstrated a strong positive relationship between the predicted and measured values in the three-dimensional space.

\begin{figure*}[!t]
	\centering
	\subfigure[]{\includegraphics[width=0.45\textwidth]{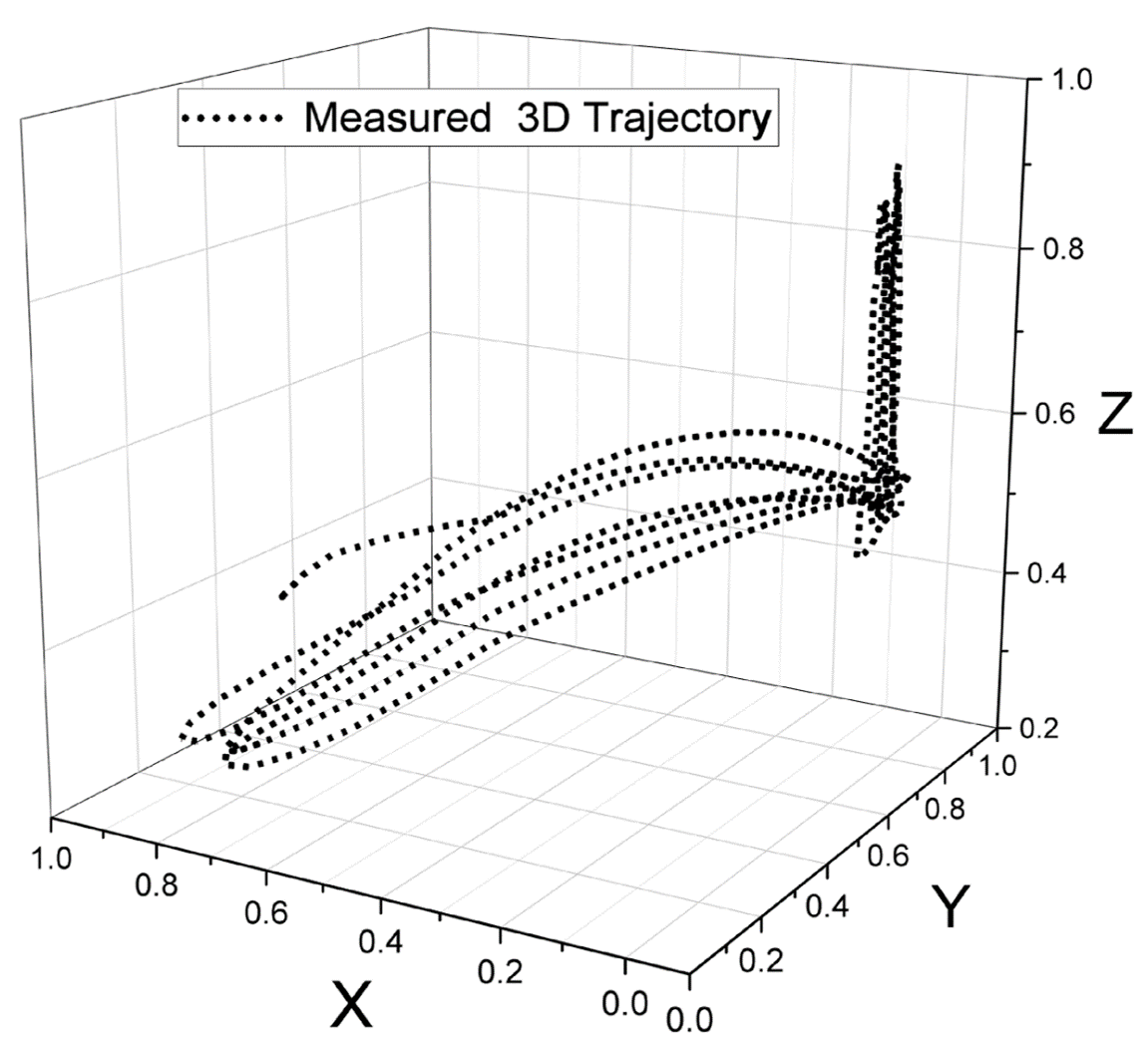}}
	\subfigure[]{\includegraphics[width=0.45\textwidth]{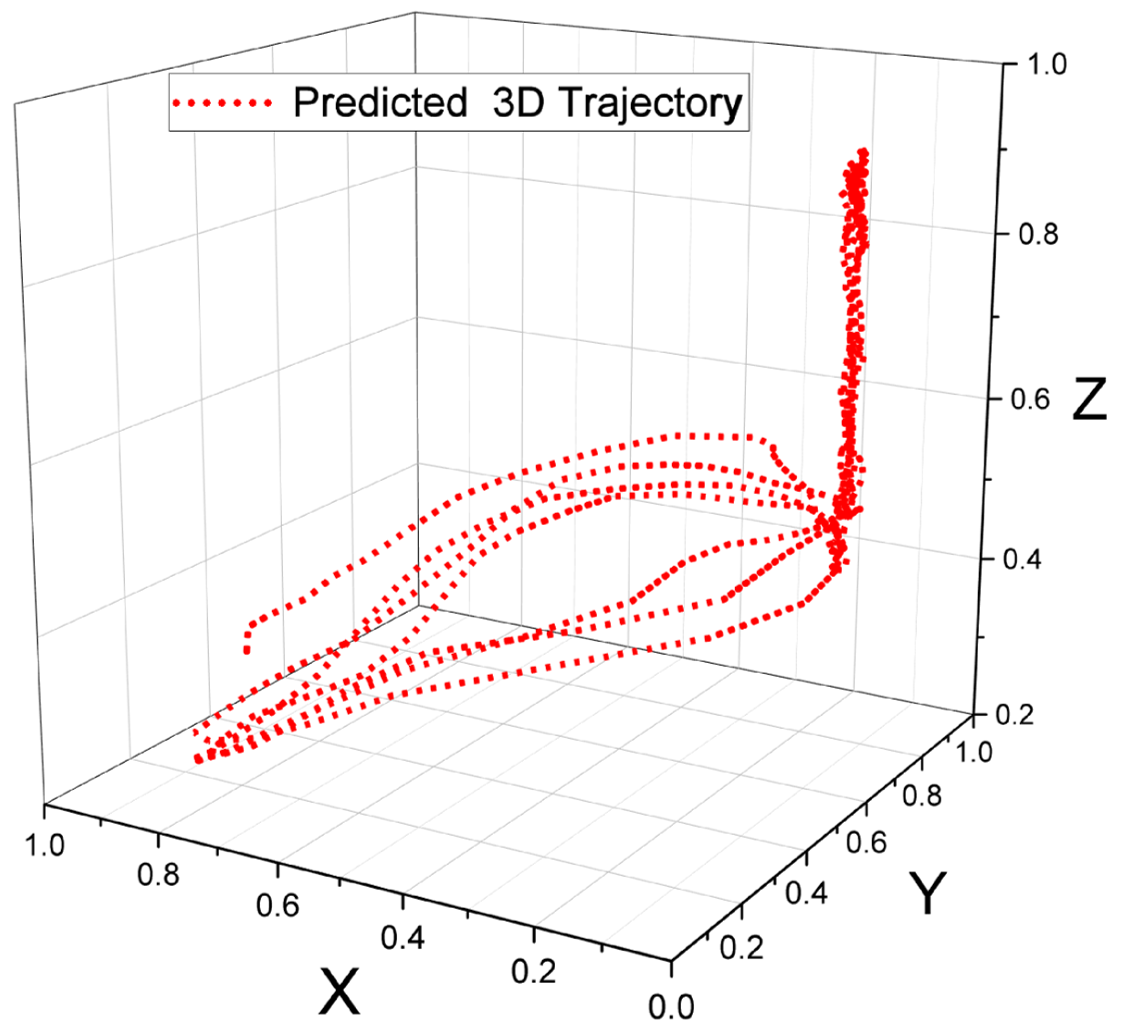}}
\caption{Comparison of Measured and Predicted Hand Kinematics in 3D space: (a) Measured: Actual hand kinematics obtained from the p4 position sensor. (b) Predicted: Hand kinematics estimated by the proposed \textit{NeuroKinect} model.}
  \label{fig:3D_result}
\end{figure*}

The $\text{MSE}_{3D}$ values further emphasize the precision and accuracy of the model's predictions. Ranging from 0.008 to 0.02, with an average $\text{MSE}_{3D}$ of 0.016 and a standard deviation of 0.001, these values indicate the model's ability to provide highly accurate predictions in the three-dimensional kinematic space. Subject P10 stood out with the lowest $\text{MSE}_{3D}$ value of 0.017, showcasing the model's exceptional precision in predicting the measured values for this particular subject. Conversely, Subject P6 had the highest $\text{MSE}_{3D}$ value of 0.021, suggesting a slightly larger deviation between the predicted and measured values. The demonstration of hand kinematics prediction in 3D space was shown in Fig. \ref{fig:3D_result}. 

One crucial aim of this research is to attain a strong correlation and close alignment between predicted and measured values. The measured and predicted 3D trajectories for subject 1 were shown in Figures 3 (a) and (b) respectively, demonstrating high correlation and low MSE, indicating accurate predictions. However, in Figures 4 (a) and (b), there is a notable deviation of the predicted values from the actual path despite the high correlation, leading to increased MSE. Ensuring high correlation and minimizing MSE is essential for achieving accurate and precise predicted values, aligning with the research's objectives.

\section{Discussion}
Our study aimed to develop a highly accurate hand kinematics prediction model using EEG signals with minimal preprocessing and a low number of epochs for training. The analysis focused on ERP polarisation and depolarization states, which provided valuable insights into the relationship between EEG signals and hand movements. Remarkably, our model successfully extracted meaningful information from the data, even with a relatively small number of trials averaging 15 epochs or less per subject during training.

The findings revealed that among the subjects in our dataset, namely subjects 1, 4, 8, 9, 10, and 11, we consistently observed significantly higher correlations compared to the other subjects in all directions. These individuals exhibited a stronger association between the recorded EEG signals and hand kinematics. Furthermore, our model's ability to achieve accurate predictions with limited samples is particularly advantageous in scenarios where data collection is challenging or time-consuming.

During our analysis, we discovered clear activation patterns in the brain during the LED stimulus and hand movement in the aforementioned subjects. These activation patterns further validated the efficacy of our model in capturing the neural correlates of hand movement and accurately predicting kinematic parameters. The consistency of these findings across multiple subjects reinforces the reliability and generalizability of our approach. Even though subjects 3 and 6 exhibited relatively weaker correlations compared to the previous group, our model's performance still surpassed that of state-of-the-art techniques. Despite the challenges posed by bad trials, where no ERP signals were associated with hand movement, we were able to extract meaningful information and achieve significant correlations in these subjects. This highlights the robustness of our model and its ability to handle challenging scenarios. For pre-processing, we applied a band-pass filter within the range of 0.5 to 12 Hz, effectively attenuating unwanted noise and isolating the relevant frequency band for analysis. We made a conscious decision not to ICA due to its computational complexity and potential hindrance to real-time implementation. This streamlined our processing pipeline and ensured the feasibility of our model for real-time applications. Our study not only uncovered significant correlations between EEG signals and hand kinematics but also delved into various analyses and explorations. These findings contribute to the advancement of the field and open up avenues for future research and development in this exciting domain. The identification of distinct activation patterns, subject-specific temporal dynamics, comparative performance evaluations, cross-validation experiments, preprocessing optimizations, and investigations into ERP components enriched our understanding of EEG-based hand kinematics prediction. The outcomes of these analyses provide a solid foundation for future research in the field of neural prostheses and pave the way for the development of more effective and efficient models.

During our analysis, we observed specific trials in which the hand kinematic estimation generated by our proposed model displayed sub-optimal results, indicating the presence of potential errors. However, the majority of the remaining trials performed well, showcasing the effectiveness of the deep learning model we employed. In order to delve deeper into the causes behind the suboptimal trials, we utilized bad trials rejection to enhance the quality of the trained model. Among the participants who took part in the study, subject P8 had the lowest number of bad trials, with only 25. Following closely behind were subjects P9 and P10, both of whom had 30 bad trials. Subject P6 had 37 bad trials, while subjects P4 and P11 had 44 bad trials each. Subject P1 had 36 bad trials, and subject P3 had the highest number of bad trials, with a total of 120. Importantly, our algorithm rejected these trials due to the absence of event-related potentials.


\section{Conclusion}

Our research introduces an innovative deep-learning model \textit{NeuroKinect} that revolutionizes BCI research by accurately estimating the hand kinematics from grasp and lift motor task. The proposed model surpasses existing techniques in accuracy and efficiency, highlighting its potential to advance BCI capabilities significantly. Leveraging a comprehensive dataset from the Grasp and Lift task, we establish robust correlations between the predicted and actual hand movements, demonstrating the remarkable capability of our model to capture the intricate relationship between EEG signals and hand kinematics. This provides a valuable tool for understanding and decoding human motor control. Furthermore, our model exhibits exceptional precision in capturing hand movements, as evidenced by the low MSE observed in the X, Y, and Z dimensions. This emphasizes its ability to reconstruct hand kinematics with high accuracy, enabling precise control and manipulation in neurorehabilitation and prosthetics applications. To enhance the integration of EEG signals, we incorporate brain source localization using the standardized sLORETA algorithm, specifically tailored for the grasp and lift movement. This additional step significantly improves the overall accuracy of our approach, unraveling the neural correlates of hand movement and enhancing the fidelity of kinematic reconstructions. An outstanding advantage of our methodology lies in its minimal preprocessing requirements, which streamline the analysis pipelines and enhance computational efficiency. This simplifies implementation and makes our model well-suited for real-time applications, ensuring timely and accurate hand kinematics reconstruction. Our model achieves excellent reconstruction performance with remarkable efficiency, converging after just 15 training epochs. This highlights its potential for real-time and adaptive applications, providing immediate and responsive feedback in scenarios that demand quick and precise motor control.

\section{Author's Contribution}
SP and AG contributed equally to this work, collaborating closely throughout the research process. SP specialized in deep learning, signal processing, and algorithm development, and was responsible for designing the architecture of the bad trail rejection algorithm, loss function, and deep learning model. AG specialized in signal processing, source localization, and temporal analysis, and conducted preprocessing and neural decoding of hand kinematics in time and space. Together, they integrated the deep learning model with the signal processing components, resulting in a successful collaboration. Both authors co-wrote the manuscript, reviewed it, and agreed to submit it for publication.

\section{Conflict of Interest Statements}
The authors declare that the research was conducted in the absence of any commercial or financial relationships that could be construed as a potential conflict of interest.

\bibliographystyle{IEEEbib}
\bibliography{Main_Article}   

\begin{thebibliography}{10}

\bibitem{makin2023neurocognitive}
Tamar~R Makin, Silvestro Micera, and Lee~E Miller,
\newblock ``Neurocognitive and motor-control challenges for the realization of
  bionic augmentation,''
\newblock {\em Nature biomedical engineering}, vol. 7, no. 4, pp. 344--348,
  2023.

\bibitem{bhagat2016design}
Nikunj~A Bhagat, Anusha Venkatakrishnan, Berdakh Abibullaev, Edward~J Artz,
  Nuray Yozbatiran, Amy~A Blank, James French, Christof Karmonik, Robert~G
  Grossman, Marcia~K O'Malley, et~al.,
\newblock ``{Design and optimization of an EEG-based brain machine interface
  (BMI) to an upper-limb exoskeleton for stroke survivors},''
\newblock {\em Frontiers in neuroscience}, vol. 10, pp. 122, 2016.

\bibitem{zhang2019eeg}
Jinhua Zhang, Baozeng Wang, Cheng Zhang, Yanqing Xiao, and Michael~Yu Wang,
\newblock ``{An EEG/EMG/EOG-based multimodal human-machine interface to
  real-time control of a soft robot hand},''
\newblock {\em Frontiers in neurorobotics}, vol. 13, pp. 7, 2019.

\bibitem{he2015wireless}
Wei He, Yue Zhao, Haoyue Tang, Changyin Sun, and Wei Fu,
\newblock ``{A wireless BCI and BMI system for wearable robots},''
\newblock {\em IEEE Transactions on Systems, Man, and Cybernetics: Systems},
  vol. 46, no. 7, pp. 936--946, 2015.

\bibitem{giri2021cortical}
Amita Giri, Lalan Kumar, and Tapan~K Gandhi,
\newblock ``Cortical source domain based motor imagery and motor execution
  framework for enhanced brain computer interface applications,''
\newblock {\em IEEE Sensors Letters}, vol. 5, no. 12, pp. 1--4, 2021.

\bibitem{gao2020classification}
Zhongke Gao, Weidong Dang, Mingxu Liu, Wei Guo, Kai Ma, and Guanrong Chen,
\newblock ``{Classification of EEG Signals on VEP-Based BCI Systems With Broad
  Learning},''
\newblock {\em IEEE Transactions on Systems, Man, and Cybernetics: Systems},
  2020.

\bibitem{li2019eeg}
Yuanqing Li, Qiyun Huang, Zhijun Zhang, Tianyou Yu, and Shenghong He,
\newblock ``{An EEG-/EOG-Based Hybrid Brain-Computer Interface: Application on
  Controlling an Integrated Wheelchair Robotic Arm System},''
\newblock {\em Frontiers in Neuroscience}, vol. 13, pp. 1243, 2019.

\bibitem{robinson2015adaptive}
Neethu Robinson, Cuntai Guan, and AP~Vinod,
\newblock ``{Adaptive estimation of hand movement trajectory in an EEG based
  brain--computer interface system},''
\newblock {\em Journal of neural engineering}, vol. 12, no. 6, pp. 066019,
  2015.

\bibitem{korik2018decoding}
Attila Korik, Ronen Sosnik, Nazmul Siddique, and Damien Coyle,
\newblock ``{Decoding imagined 3D hand movement trajectories from EEG: evidence
  to support the use of mu, beta, and low gamma oscillations},''
\newblock {\em Frontiers in neuroscience}, vol. 12, pp. 130, 2018.

\bibitem{sosnik2020reconstruction}
Ronen Sosnik and Omer~Ben Zur,
\newblock ``{Reconstruction of hand, elbow and shoulder actual and imagined
  trajectories in 3D space using EEG slow cortical potentials},''
\newblock {\em Journal of Neural Engineering}, vol. 17, no. 1, pp. 016065,
  2020.

\bibitem{chen2022continuous}
Yi-Feng Chen, Ruiqi Fu, Junde Wu, Jongbin Song, Rui Ma, Yi-Chuan Jiang, and
  Mingming Zhang,
\newblock ``Continuous bimanual trajectory decoding of coordinated movement
  from eeg signals,''
\newblock {\em IEEE Journal of Biomedical and Health Informatics}, vol. 26, no.
  12, pp. 6012--6023, 2022.

\bibitem{degenhart2020stabilization}
Alan~D Degenhart, William~E Bishop, Emily~R Oby, Elizabeth~C Tyler-Kabara,
  Steven~M Chase, Aaron~P Batista, and Byron~M Yu,
\newblock ``Stabilization of a brain-computer interface via the alignment of
  low-dimensional spaces of neural activity,''
\newblock {\em Nature biomedical engineering}, vol. 4, no. 7, pp. 672--685,
  2020.

\bibitem{grech2008review}
Roberta Grech, Tracey Cassar, Joseph Muscat, Kenneth~P Camilleri, Simon~G
  Fabri, Michalis Zervakis, Petros Xanthopoulos, Vangelis Sakkalis, and Bart
  Vanrumste,
\newblock ``{Review on solving the inverse problem in EEG source analysis},''
\newblock {\em Journal of neuroengineering and rehabilitation}, vol. 5, no. 1,
  pp. 25, 2008.

\bibitem{akaike1974new}
Hirotugu Akaike,
\newblock ``{A new look at the statistical model identification},''
\newblock {\em IEEE transactions on automatic control}, vol. 19, no. 6, pp.
  716--723, 1974.

\bibitem{wax1985detection}
Mati Wax and Thomas Kailath,
\newblock ``{Detection of signals by information theoretic criteria},''
\newblock {\em IEEE Transactions on acoustics, speech, and signal processing},
  vol. 33, no. 2, pp. 387--392, 1985.

\bibitem{green1988transformation}
Andrew~A Green, Mark Berman, Paul Switzer, and Maurice~D Craig,
\newblock ``{A transformation for ordering multispectral data in terms of image
  quality with implications for noise removal},''
\newblock {\em IEEE Transactions on geoscience and remote sensing}, vol. 26,
  no. 1, pp. 65--74, 1988.

\bibitem{giri2023f}
Amita Giri, John~C Mosher, Amir Adler, and Dimitrios Pantazis,
\newblock ``An f-ratio-based method for estimating the number of active sources
  in meg,''
\newblock {\em arXiv preprint arXiv:2306.05892}, 2023.

\bibitem{mosher_multiple_1992}
J.C. Mosher, P.S. Lewis, and R.M. Leahy,
\newblock ``Multiple dipole modeling and localization from spatio-temporal
  {MEG} data,''
\newblock {\em IEEE Transactions on Biomedical Engineering}, vol. 39, no. 6,
  pp. 541--557, June 1992.

\bibitem{Mosher1999}
J.~C. {Mosher} and R.~M. {Leahy},
\newblock ``Source localization using recursively applied and projected {(RAP)
  MUSIC},''
\newblock {\em IEEE Transactions on Signal Processing}, vol. 47, no. 2, pp.
  332--340, Feb 1999.

\bibitem{makela_truncated_2018}
N.~Mäkelä, M.~Stenroos, J.~Sarvas, and R.~J. Ilmoniemi,
\newblock ``Truncated {RAP}-{MUSIC} ({TRAP}-{MUSIC}) for {MEG} and {EEG} source
  localization,''
\newblock {\em NeuroImage}, vol. 167, pp. 73--83, Feb. 2018.

\bibitem{ilmoniemi2019brain}
R.J. Ilmoniemi and J.~Sarvas,
\newblock {\em Brain Signals: Physics and Mathematics of MEG and EEG},
\newblock MIT Press, 2019.

\bibitem{giri2018eeg}
Amita Giri, Lalan Kumar, and Tapan Gandhi,
\newblock ``{EEG dipole source localization in hemispherical harmonics
  domain},''
\newblock in {\em 2018 Asia-Pacific Signal and Information Processing
  Association Annual Summit and Conference (APSIPA ASC)}. IEEE, 2018, pp.
  679--684.

\bibitem{giri2019head}
Amita Giri, Lalan Kumar, and Tapan Gandhi,
\newblock ``{Head Harmonics Based EEG Dipole Source Localization},''
\newblock in {\em 2019 53rd Asilomar Conference on Signals, Systems, and
  Computers}. IEEE, 2019, pp. 2149--2153.

\bibitem{giri2020brain}
Amita Giri, Lalan Kumar, and Tapan~Kumar Gandhi,
\newblock ``{Brain source localization in head harmonics domain},''
\newblock {\em IEEE Transactions on Instrumentation and Measurement}, vol. 70,
  pp. 1--10, 2020.

\bibitem{giri2022anatomical}
Amita Giri, Lalan Kumar, Nilesh Kurwale, and Tapan~K Gandhi,
\newblock ``{Anatomical harmonics basis based brain source localization with
  application to epilepsy},''
\newblock {\em Scientific Reports}, vol. 12, no. 1, pp. 11240, 2022.

\bibitem{hamalainen1994interpreting}
Matti~S H{\"a}m{\"a}l{\"a}inen and Risto~J Ilmoniemi,
\newblock ``Interpreting magnetic fields of the brain: minimum norm
  estimates,''
\newblock {\em Medical \& biological engineering \& computing}, vol. 32, pp.
  35--42, 1994.

\bibitem{DSPM}
{A. M. Dale, A. K. Liu, B. R. Fischl, R. L. Buckner, J. W. Belliveau, J. D.
  Lewine and E. Halgren},
\newblock ``Dynamic statistical parametric mapping: Combining {fMRI} and {MEG}
  for high-resolution imaging of cortical activity,''
\newblock {\em Neuron}, vol. 26, no. 1, pp. 55--67, Apr 2000.

\bibitem{pascual2002standardized}
Roberto~Domingo Pascual-Marqui et~al.,
\newblock ``{Standardized low-resolution brain electromagnetic tomography
  (sLORETA): technical details},''
\newblock {\em Methods Find Exp Clin Pharmacol}, vol. 24, no. Suppl D, pp.
  5--12, 2002.

\bibitem{pancholi2019improved}
Sidharth Pancholi and Amit~M Joshi,
\newblock ``Improved classification scheme using fused wavelet packet transform
  based features for intelligent myoelectric prostheses,''
\newblock {\em IEEE Transactions on Industrial Electronics}, vol. 67, no. 10,
  pp. 8517--8525, 2019.

\bibitem{pancholi2019electromyography}
Sidharth Pancholi and Amit~M Joshi,
\newblock ``Electromyography-based hand gesture recognition system for upper
  limb amputees,''
\newblock {\em IEEE Sensors Letters}, vol. 3, no. 3, pp. 1--4, 2019.

\bibitem{pancholi2022dlpr}
Sidharth Pancholi, Amit~M Joshi, and Deepak Joshi,
\newblock ``Dlpr: Deep learning-based enhanced pattern recognition frame-work
  for improved myoelectric prosthesis control,''
\newblock {\em IEEE Transactions on Medical Robotics and Bionics}, vol. 4, no.
  4, pp. 991--999, 2022.

\bibitem{bang2021spatio}
Ji-Seon Bang, Min-Ho Lee, Siamac Fazli, Cuntai Guan, and Seong-Whan Lee,
\newblock ``Spatio-spectral feature representation for motor imagery
  classification using convolutional neural networks,''
\newblock {\em IEEE Transactions on Neural Networks and Learning Systems}, vol.
  33, no. 7, pp. 3038--3049, 2021.

\bibitem{jeong2020brain}
Ji-Hoon Jeong, Kyung-Hwan Shim, Dong-Joo Kim, and Seong-Whan Lee,
\newblock ``Brain-controlled robotic arm system based on multi-directional
  cnn-bilstm network using eeg signals,''
\newblock {\em IEEE Transactions on Neural Systems and Rehabilitation
  Engineering}, vol. 28, no. 5, pp. 1226--1238, 2020.

\bibitem{yin2017cross}
Zhong Yin and Jianhua Zhang,
\newblock ``Cross-subject recognition of operator functional states via eeg and
  switching deep belief networks with adaptive weights,''
\newblock {\em Neurocomputing}, vol. 260, pp. 349--366, 2017.

\bibitem{zhang2023vit}
Hanyang Zhang, Ke~Yang, Gangsheng Cao, and Chunming Xia,
\newblock ``Vit-llmr: Vision transformer-based lower limb motion recognition
  from fusion signals of mmg and imu,''
\newblock {\em Biomedical Signal Processing and Control}, vol. 82, pp. 104508,
  2023.

\bibitem{zhang2020expression}
Hongli Zhang,
\newblock ``Expression-eeg based collaborative multimodal emotion recognition
  using deep autoencoder,''
\newblock {\em IEEE Access}, vol. 8, pp. 164130--164143, 2020.

\bibitem{ju2022tensor}
Ce~Ju and Cuntai Guan,
\newblock ``Tensor-cspnet: A novel geometric deep learning framework for motor
  imagery classification,''
\newblock {\em IEEE Transactions on Neural Networks and Learning Systems},
  2022.

\bibitem{gong2021deep}
Shu Gong, Kaibo Xing, Andrzej Cichocki, and Junhua Li,
\newblock ``Deep learning in eeg: Advance of the last ten-year critical
  period,''
\newblock {\em IEEE Transactions on Cognitive and Developmental Systems}, vol.
  14, no. 2, pp. 348--365, 2021.

\bibitem{khare2020time}
Smith~K Khare and Varun Bajaj,
\newblock ``Time--frequency representation and convolutional neural
  network-based emotion recognition,''
\newblock {\em IEEE transactions on neural networks and learning systems}, vol.
  32, no. 7, pp. 2901--2909, 2020.

\bibitem{tiwari2022midnn}
Smita Tiwari, Shivani Goel, and Arpit Bhardwaj,
\newblock ``Midnn-a classification approach for the eeg based motor imagery
  tasks using deep neural network,''
\newblock {\em Applied Intelligence}, pp. 1--20, 2022.

\bibitem{cho2021neurograsp}
Jeong-Hyun Cho, Ji-Hoon Jeong, and Seong-Whan Lee,
\newblock ``Neurograsp: Real-time eeg classification of high-level motor
  imagery tasks using a dual-stage deep learning framework,''
\newblock {\em IEEE Transactions on Cybernetics}, vol. 52, no. 12, pp.
  13279--13292, 2021.

\bibitem{luciw2014multi}
Matthew~D Luciw, Ewa Jarocka, and Benoni~B Edin,
\newblock ``{Multi-channel EEG recordings during 3,936 grasp and lift trials
  with varying weight and friction},''
\newblock {\em Scientific data}, vol. 1, no. 1, pp. 1--11, 2014.

\bibitem{szegedy2015going}
Christian Szegedy, Wei Liu, Yangqing Jia, Pierre Sermanet, Scott Reed, Dragomir
  Anguelov, Dumitru Erhan, Vincent Vanhoucke, and Andrew Rabinovich,
\newblock ``Going deeper with convolutions,''
\newblock in {\em Proceedings of the IEEE conference on computer vision and
  pattern recognition}, 2015, pp. 1--9.

\bibitem{totaro2020non}
Simone Totaro, Amir Hussain, and Simone Scardapane,
\newblock ``A non-parametric softmax for improving neural attention in
  time-series forecasting,''
\newblock {\em Neurocomputing}, vol. 381, pp. 177--185, 2020.

\bibitem{pancholi2022source}
Sidharth Pancholi, Amita Giri, Anant Jain, Lalan Kumar, and Sitikantha Roy,
\newblock ``{Source aware deep learning framework for hand kinematic
  reconstruction using EEG signal},''
\newblock {\em IEEE Transactions on Cybernetics}, 2022.

\end{thebibliography}

\end{document}